\documentclass[aps,groupedaddress,preprint,superscriptaddress,showpacs]{revtex4-1}
\usepackage{eurosym}
\usepackage{amsmath}
\usepackage{fixmath}
\usepackage{graphicx}
\usepackage{wrapfig}
\usepackage[toc,page]{appendix}
\usepackage{subcaption}

\usepackage{hyperref}
\usepackage[capitalise]{cleveref}
\usepackage{subcaption}

\usepackage{amsfonts}

\usepackage{xcolor}

\graphicspath{{images/}}

\newcommand{\be}{\begin{equation}}
\newcommand{\ee}{\end{equation}}

\begin{document}

\title{Dynamics and control of entangled electron-photon states in  nanophotonic systems with time-variable parameters}

\author{Qianfan Chen}
\affiliation{Department of Physics and Astronomy, Texas A\&M University, College Station, TX, 77843 USA}
\author{Yongrui Wang}
\affiliation{Department of Physics and Astronomy, Texas A\&M University, College Station, TX, 77843 USA}
\author{Sultan Almutairi}
\affiliation{Department of Physics and Astronomy, Texas A\&M University, College Station, TX, 77843 USA}
\author{ Maria Erukhimova}
\affiliation{Institute of Applied Physics, Russian Academy of Sciences, Nizhny Novgorod, 603950, Russia }
\author{ Mikhail Tokman}
\affiliation{Institute of Applied Physics, Russian Academy of Sciences, Nizhny Novgorod, 603950, Russia }
\author{Alexey Belyanin}
\affiliation{Department of Physics and Astronomy, Texas A\&M University, College Station, TX, 77843 USA}

\begin{abstract}

We study the dynamics of strongly coupled nanophotonic systems with time-variable parameters. The approximate analytic solutions are obtained for a broad class of open quantum systems including a two-level fermion emitter strongly coupled to a multimode quantized electromagnetic field in a cavity with time-varying cavity resonances or the electron transition energy. The coupling of the fermion and photon subsystems to their dissipative reservoirs is included within the stochastic equation of evolution approach, which is equivalent to the Lindblad approximation in the master equation formalism. The analytic solutions for the quantum states and the observables are obtained under the approximation that the rate of parameter modulation and the amplitude of the frequency modulation are much smaller than the optical transition frequencies. At the same time, they can be arbitrary with respect to the generalized Rabi oscillations frequency which determines the coherent dynamics. Therefore, our analytic theory can be applied to an arbitrary modulation of the parameters, both slower and faster than the Rabi frequency, for complete control of the quantum state. In particular, we demonstrate protocols for switching on and off the entanglement between the fermionic and photonic degrees of freedom, swapping between the quantum states, and the decoupling of the fermionic qubit from the cavity field due to modulation-induced transparency.

\end{abstract}

\date{\today }

\maketitle

\section{Introduction}

Solid-state photonic qubits based on the fermion
systems coupled  to a quantized electromagnetic (EM) field in a plasmonic or dielectric nanocavity are promising for a variety of quantum information and quantum sensing applications \cite{haroche,lodahl2015,degen2017}.  Their benefits include compatibility with semiconductor technology, scalability, and potential for operation at temperatures much higher than the alternative platforms based on superconducting qubits or trapped ions. Indeed, strong coupling to single quantum emitters in dielectric nanocavities was demonstrated in various systems, for example color centers \cite{lukin2016} or quantum dots (QDs) Refs.~\cite{deppe,reithmaier}. In plasmonic cavities, strong coupling to single molecules  Refs.~\cite{chikkaraddy2016,benz2016, park2016} and colloidal QDs \cite{pelton2018, gross2018, park2019} has been achieved at room temperature; see, e.g., Refs.~\cite{lodahl2015,thorma2015,bitton2019,kono2019,may2020} for recent reviews. 

While the quantum dynamics of entangled nanophotonic systems is interesting by itself, many applications would benefit from to control and modify the qubit states by time-dependent variation of certain parameters, while taking into account various processes of decoherence and dissipation. There is of course a large body of work related to cavity quantum electrodynamics (QED) with time-variable parameters. For example, the dynamics of nanophotonics systems with periodic modulation of some parameter, such as the cavity size or the position of a quantum emitter in a cavity, has been studied extensively in the burgeoning fields of cavity optomechanics \cite{aspelmeyer2014,meystre2013,pirkkalainen2015} and quantum acoustics \cite{chu2017,hong2017, arriola2019}. In this case the most interesting new element added to the nanophotonic system is the parametric resonance or the dressing of the electron-photon coupling by mechanical oscillations. Near the parametric resonance, the system can be mapped to an exactly solvable time-independent Hamiltonian  within the rotating-wave approximation  \cite{tokman2020}.  

There is a class of time-dependent Hamiltonians for which the nonstationary Schr\"{o}dinger equation can be solved exactly in the analytic form, notably multistate Landau-Zener Hamiltonians and driven Tavis-Cummings Hamiltonians \cite{sinitsyn2018, chernyak2018}; see also \cite{li2018} where this technique was applied to the quantum annealing problem. Here we are interested in the nanophotonic applications, so we have to consider open multimode photonic systems with an arbitrary time dependence of the parameters. Therefore, we restrict ourselves to the adiabatic dynamics, for which the analytic solution can be found for a broad variety of systems with time-dependent cavity or fermion emitter parameters, and with dissipation included at the level of the Lindblad formalism. We find that the condition of adiabaticity is not that restrictive; in particular it still allows one to consider the parameter variation at a rate comparable to or faster than the generalized Rabi frequency in strongly coupled systems, which may be required for qubit manipulation. 

We will also stick to the rotating wave
approximation (RWA)  \cite{Scully1997}.
The use of RWA restricts the coupling strength to the values much lower than
the characteristic energies in the system, such as the optical transition or
photon energy. The emerging studies of the so-called ultra-strong
coupling regime \cite{kono2019} have to go beyond the RWA. Nevertheless, for the
vast majority of experiments, including nonperturbative strong coupling
dynamics and entanglement, the RWA is adequate and provides some crucial
simplifications that allow one to obtain analytic solutions.

In particular, within Schr\"{o}dinger's description, the equations of motion
for the components of an infinitely dimensional state vector $\left\vert
\Psi \right\rangle $ that describes a coupled fermion-boson system can be
split into the blocks of low dimensions if the RWA is applied. This is true
even if the dynamics of the fermion subsystem is nonperturbative, e.g. the
effects of saturation are important. Note that there is no such
simplification in the Heisenberg representation, except within the perturbation theory; see e.g. \cite{Scully1997}. This is  because boson operators are defined on a basis of infinite dimension and truncation of their dynamics into blocks of small dimensions is generally not possible (see also \cite{tokman2020}).  The Schr\"{o}dinger's approach also leads to fewer equations
for the state vector components than the approach based on the von Neumann
master equation for the elements of the density matrix. This is especially true for a system with many degrees of freedom, e.g., many electron states coupled to multiple boson field modes. 

Obviously, the Schr\"{o}dinger equation in its standard form cannot be
applied to describe open systems coupled to a dissipative reservoir. In this
case the stochastic versions of the equation of evolution for the state vector
have been developed, e.g. the method of quantum jumps \cite{Scully1997,Plenio1998}. This method is optimal for numerical analysis in
the Monte-Carlo type schemes. Here we formulate the stochastic equation which is more
conducive to the analytic treatment. In \cite{tokman2020} we showed that the stochastic equation of evolution for the state vector can be derived directly from the Heisenberg-Langevin formalism. 

The paper is structured as follows. Section II formulates the model and the Hamiltonian for two-level electron system and a quantized EM field in a nanocavity with time-variable parameters. It treats a single-mode cavity in detail as a particular case and describes simple manipulations with a single cavity mode coupled to a single fermionic qubit. Section III considers the dynamics of two time-modulated cavity modes coupled to a single quantum emitter and Sec.~IV treats the case of a variable frequency of the optical transition in a fermion qubit.  Section V solves the quantum dynamics for an open time-dependent system with the coupling to dissipative reservoirs taken into account. An interesting phenomenon of modulation-induced transparency is analyzed. Numerical estimations for various nanophotonic systems reported in the literature are presented. Conclusions are in Section VI. Appendix A describes the quantization procedure for a plasmon cavity field with strongly subwavelength localization.  Appendix B summarizes the main properties of the stochastic equation of evolution and compares with the Lindblad density-matrix formalism.

\section{Cavity QED with time-variable parameters}

\subsection{Standard cavity QED Hamiltonian for a quantized field coupled to a two-level emitter}

For reference, we start from summarizing basic textbook  facts about a quantized electron system resonantly coupled to the quantum multimode EM field of a
nanocavity without any time dependence, and then consider the time-dependent models in the next sections.   

%%%%%%%%%%%%%%%%%%%%%%%%%%%%%%%

\begin{figure}[htb]
\centering

\begin{subfigure}[b]{0.6\textwidth}
\includegraphics[width=1\linewidth]{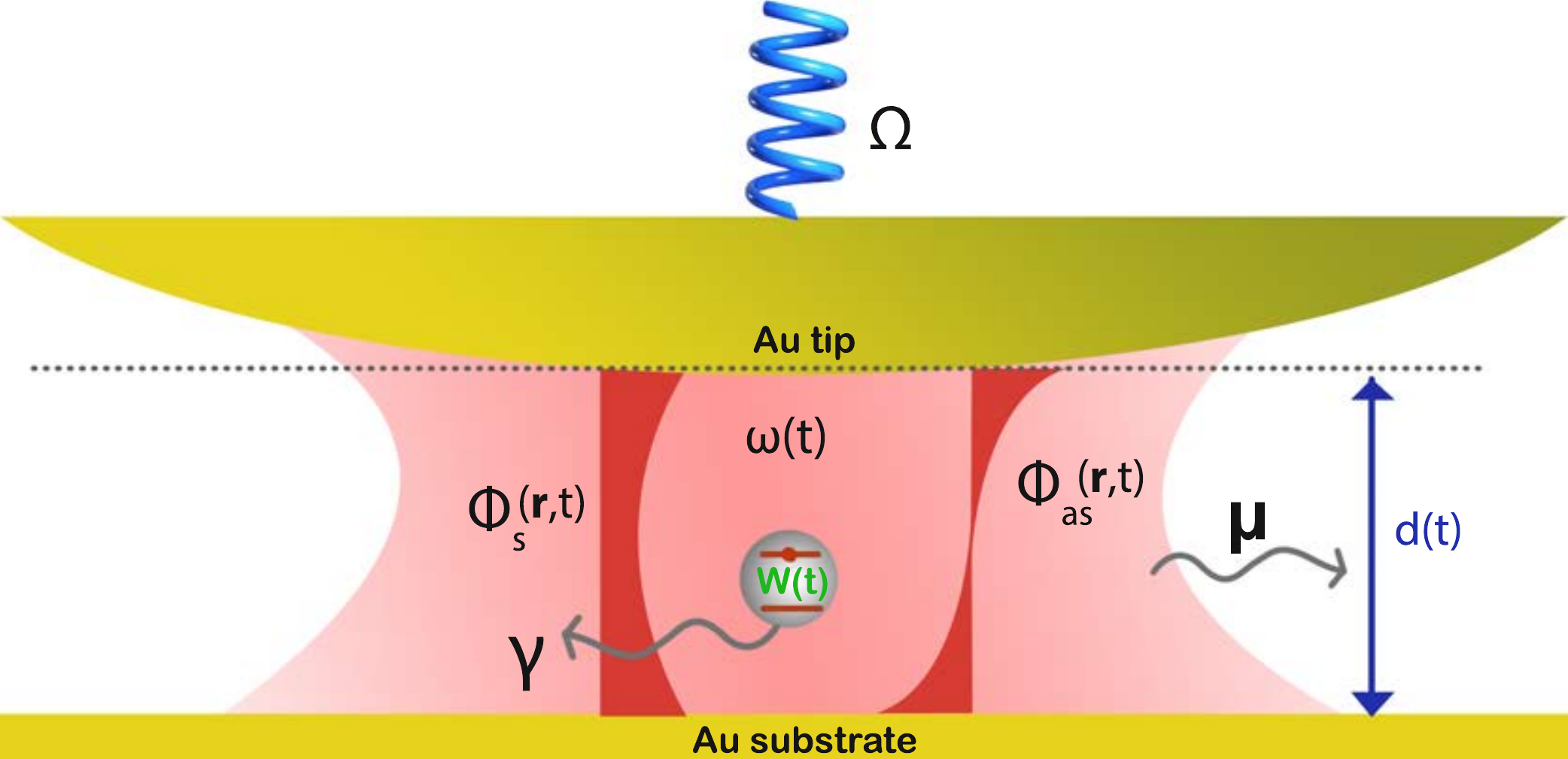}
\caption{ }
\label{fig1}
\end{subfigure}

\begin{subfigure}[b]{0.6\textwidth}
\includegraphics[width=1\linewidth]{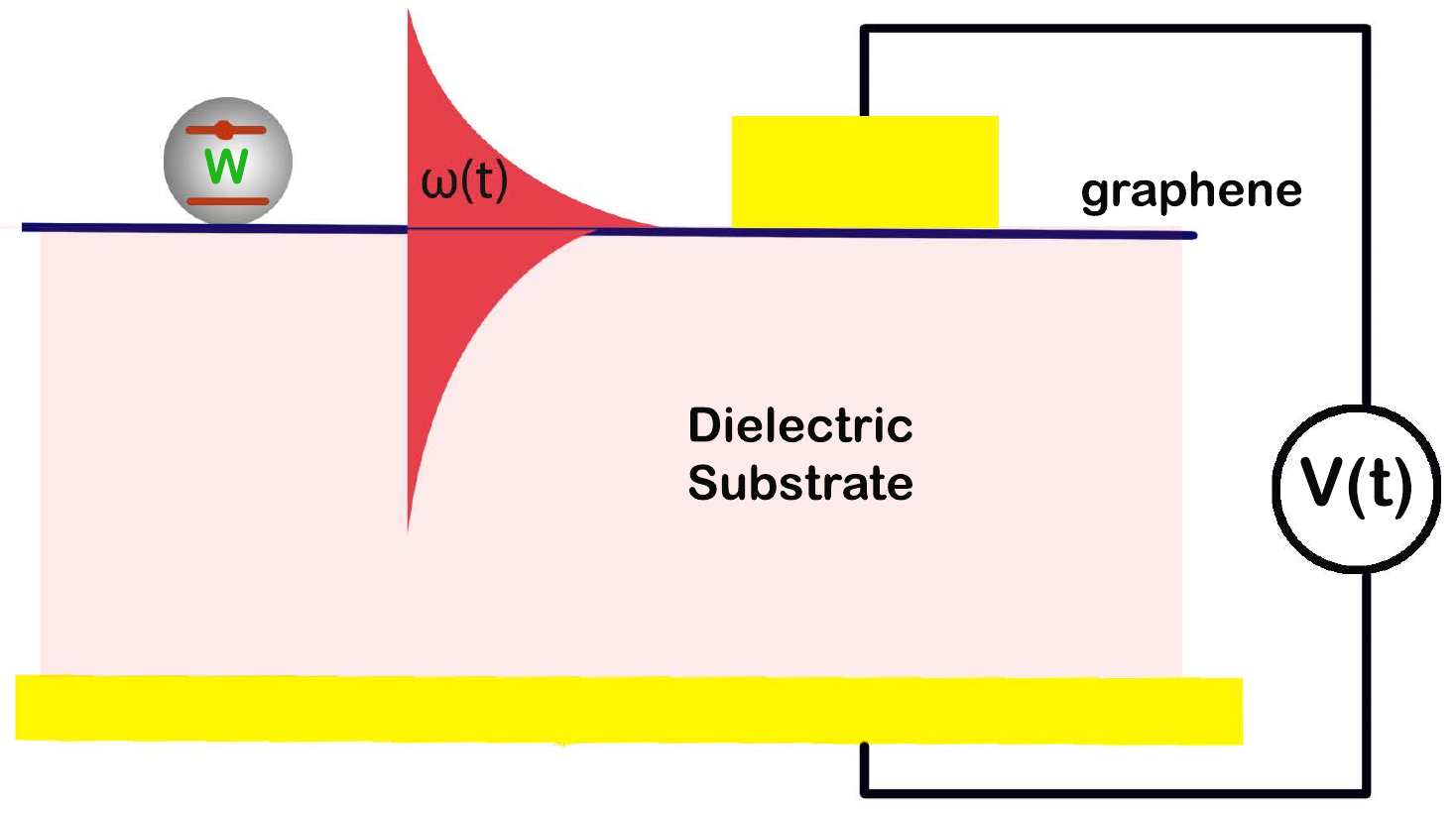}
\caption{}
\label{fig1}
\end{subfigure}

\caption { (a) A sketch of a quantum emitter (e.g. a quantum dot or a single molecule) in a nanocavity with time-dependent parameters created by
a metallic nanotip of the scanning probe and a metallic substrate. The profiles of the electric potential $ \Phi\left( \mathbf{r}, t\right)$ for the symmetric and antisymmetric mode (see Appendix A) are sketched. Other parameters are the transition energy $W(t)$ for a quantum emitter, the optical field frequency $\omega(t)$, the cavity height $d(t)$, and the relaxation constants of the cavity field, $\mu$, and a quantum emitter, $\gamma$.  (b) A quantum emitter coupled to the cavity surface plasmon field supported by graphene. The mode frequency $\omega(t)$ can be varied by applying variable voltage $V(t)$ which modifies the charge density in graphene. }

\end{figure}

%%%%%%%%%%%%%%%%%%%%%%%%%

Consider the simplest version of the fermion subsystem: two electron states $
\left\vert 0\right\rangle $ and $\left\vert 1\right\rangle $ with energies $
0 $ and $W$, respectively. We will call it an 
``atom'' for brevity, although it can be electron states of a molecule, a
quantum dot, a defect in a semiconductor, or any other electron system. Introduce creation and
annihilation operators of the excited state $\left\vert 1\right\rangle $, 
$
\hat{\sigma}=\left\vert 0\right\rangle \left\langle 1\right\vert$, $\hat{\sigma 
}^{\dagger }=\left\vert 1\right\rangle \left\langle 0\right\vert$,
which satisfy standard commutation relations for fermions:
\begin{equation*}
\hat{\sigma}^{\dagger }\left\vert 0\right\rangle =\left\vert 1\right\rangle ,%
\hat{\sigma}\left\vert 1\right\rangle =\left\vert 0\right\rangle ,\hat{\sigma%
}\hat{\sigma}=\hat{\sigma}^{\dagger }\hat{\sigma}^{\dagger }=0;\left[ \hat{%
\sigma},\hat{\sigma}^{\dagger }\right] _{+}=\hat{\sigma}\hat{\sigma}%
^{\dagger }+\hat{\sigma}^{\dagger }\hat{\sigma}=1.
\end{equation*}%
The Hamiltonian of an atom is 
\begin{equation}
\hat{H}_{a}=W\hat{\sigma}^{\dagger }\hat{\sigma}.  \label{ha}
\end{equation}%
We will also need the dipole moment operator, $
\mathbf{\hat{d}}=\mathbf{d}\left( \hat{\sigma}^{\dagger }+\hat{\sigma}%
\right)$, 
where $\mathbf{d=}\left\langle 1\right\vert \mathbf{\hat{d}}\left\vert
0\right\rangle $ is a real vector. For a finite motion we can always choose
the coordinate representation of stationary states in terms of real
functions.

We assume that an atom is placed in a nanocavity and is resonantly coupled to the electric field of quantized cavity modes. Figure 1 sketches two out of many possible geometries of a time-variable nanocavity, e.g. formed by the nanotip of the scanning probe and the metallic substrate (Fig.~1a), similar to the recent experiments with strong coupling to single quantum emitters \cite{park2016, pelton2018, gross2018, park2019}. Of course many other cavity geometries are possible, such as the one in Fig.~1b where the quantum emitter is coupled to the cavity surface plasmon field supported by graphene.  Here the optical transition energy $W(t)$, the photon mode frequency $\omega(t)$, and field amplitudes described by an electric potential $ \Phi\left( \mathbf{r}, t\right)$  are all subject to external modulation by e.g.~varying the tip distance to the substrate, the position of a quantum emitter in a cavity, or a variable voltage applied to graphene or to a QD in a semiconductor nanostructure, but we will start from the Hamiltonian without any time dependence for future comparison.

We use a standard representation for the electric field operator in a cavity:%
\begin{equation}
\mathbf{\hat{E}}=\sum_{i}\left[ \mathbf{E}_{i}\left( \mathbf{r}\right) \hat{c%
}_{i}+\mathbf{E}_{i}^{\ast }\left( \mathbf{r}\right) \hat{c}_{i}^{\dagger }%
\right] ,  \label{ef}
\end{equation}%
where $\hat{c}_{i}^{\dagger },\hat{c}_{i}$ are creation and annihilation
operators for photons at frequency $\omega _{i}$; the functions $%
\mathbf{E}_{i}\left( \mathbf{r}\right) $ describe the spatial structure of
the EM modes in a cavity. The relation between the modal frequency $\omega
_{i}$ and the function $\mathbf{E}_{i}\left( \mathbf{r}\right) $ can
be found by solving the boundary-value problem of the classical
electrodynamics \cite{Scully1997}. The normalization conditions \cite{Tokman2016}
\begin{equation}
\int_{V}\frac{\partial \left[ \omega_i^{2}\varepsilon \left( \omega_i,\mathbf{r%
}\right) \right] }{\omega_i \partial \omega_i }\mathbf{E}_{i}^{\ast }\left(
\mathbf{r}\right) \mathbf{E}_{i}\left( \mathbf{r}\right) d^{3}r=4\pi \hbar
\omega _{i}  \label{nc-const}
\end{equation}%
ensure correct bosonic commutators $\left[ \hat{c}_{i},\hat{c}_{i}^{\dagger }%
\right] =\delta _{ij}$ and the field Hamiltonian in the form
\begin{equation}
\hat{H}_{em}=\hbar \sum_{i}\omega _{i}\left( \hat{c}_{i}^{\dagger }\hat{c}%
_{i}+\frac{1}{2}\right) .  \label{hem}
\end{equation}%
Here $V$ is a quantization volume  and $\varepsilon \left( \omega ,\mathbf{r}%
\right) $ is the dielectric function of a dispersive medium that fills the
cavity. 

 Equation (\ref{nc-const}) is true for any fields satisfying Maxwell's equations as long as intracavity losses can be neglected and the flux of the Poynting vector through the total cavity surface is zero; see, e.g., Refs.~\cite{Tokman2016,Tokman2013, Tokman2015, Tokman2018}. Of course the photon losses are always important when calculating the decoherence rates and fluctuations. What matters for Eq.~(\ref{nc-const}) is that the effect of losses on the {\it spatial structure} of the cavity modes is insignificant. The latter is true as long as it makes sense to talk about cavity modes at all, which means in practice that the cavity Q-factor is at least around 10 or greater. 

In many experiments involving strong coupling to a single quantum emitter the plasmonic cavities of nanometer size and even below 1 nm are used. The quantization procedure for a strongly subwavelength plasmon field has its peculiarities. We describe it in detail in Appendix A.

Adding
the interaction Hamiltonian with a EM cavity mode in the electric dipole
approximation, $-\mathbf{\hat{d}}\cdot \mathbf{\hat{E}}$, the Hamiltonian of
an atom coupled to a single mode EM field is 
\begin{equation}
\hat{H}=\hat{H}_{em}+\hat{H}_{a}-\mathbf{d}\left( \hat{\sigma}^{\dagger }+%
\hat{\sigma}\right) \cdot \left[ \mathbf{E}\left( \mathbf{r}\right) \hat{c}+%
\mathbf{E}^{\ast }\left( \mathbf{r}\right) \hat{c}^{\dagger }\right] _{%
\mathbf{r}=\mathbf{r}_{a}},  \label{h}
\end{equation}%
where $\mathbf{r}=\mathbf{r}_{a}$ denotes the position of an atom inside the
cavity. This can be rewritten as 
\begin{eqnarray}
\hat{H} &=&\hat{H}_{em}+\hat{H}_{a} - \left( \chi \hat{\sigma}
^{\dagger }\hat{c}+\chi ^{\ast }\hat{\sigma}\hat{c}^{\dagger }+\chi \hat{
\sigma}\hat{c}+\chi ^{\ast }\hat{\sigma}^{\dagger }\hat{c}^{\dagger }\right)
 \label{toh}
\end{eqnarray}
where
$ \chi =\left( \mathbf{d}\cdot \mathbf{E}\right) _{\mathbf{r}=
\mathbf{r}_{a}}$. 

The best conditions for entanglement are realized in the vicinity of an atom-field resonance, where one can apply the rotating wave approximation (RWA). The RWA Hamiltonian is obtained by dropping the last two terms in Eq.~(\ref{toh}). 
Note that we can always take the functions $\mathbf{E}\left( \mathbf{r}%
\right) $ to be real at
the position of an atom. This single-mode model is of course the Jaynes-Cummings Hamiltonian \cite{JC1963}.

%%%%%%%%%%%%%%%%%%%%%%%%%%%%%%%%%%%%%%

\subsection{Quantized electromagnetic field in a time-variable cavity}

In a standard approach to quantization of the EM field based on Eqs.~(\ref{ef})-(\ref{hem}), a set of mode
frequencies $\omega _{i}$ and the relation between the frequency $%
\omega _{i}$ and the spatial structure of the field mode $\mathbf{E} 
_{i}\left( \mathbf{r}\right) $ are determined by solving the boundary-value
problem for the classical EM field. Let's assume that the solution of this
boundary-value problem depends on a certain parameter $p$, for example the
cavity height $d(t)$ in Fig.~1 or the position of the emitter with respect to the field
distribution. In this case $\omega _i(p)$ and $%
\mathbf{E}_{i}\left( \mathbf{r,}p\right) $ are functions of $p$. Of course
the solution depends on many parameters of the cavity, but we consider the
situation when this particular parameter is adiabatically changing with
time. As usual, ``adiabatically'' means that the change can be
arbitrary (e.g. periodic or not) but it should be slow as compared to typical frequencies of all subsystems when the parameters are constant, such as 
the modal frequencies and the transition frequency of a quantum emitter. It is important that the rate of change of parameters can be arbitrary as compared to characteristic frequency scales which determine the interaction between subsystems, such as the Rabi frequency, as long as these scales are smaller than the modal or transition frequencies \cite{kruskal1962,tokman2011}. 

For an adiabatically varying parameter Eqs.~(\ref{ef})-(\ref{hem}) depend on the
instantaneous value of the parameter,
\begin{equation}
\mathbf{\hat{E}}=\sum_{i}\left[ \mathbf{E}_{i}\left( \mathbf{r,}p\right)
\hat{c}_{i}+\mathbf{E}_{i}^{\ast }\left( \mathbf{r,}p\right) \hat{c}%
_{i}^{\dagger }\right] ,  \label{elef}
\end{equation}
\begin{equation}
\hat{H}\ =\sum_{i}\hat{H}_{i}\ \ \ \hat{H}_{i}=\hbar \omega _i(p) \left( \hat{c}_{i}^{\dagger }\hat{c}_{i}+\frac{1}{2}\right)
,   \label{Hem}
\end{equation}
\begin{equation}
\int_{V\left( p\right) }\frac{\partial \left[ \omega _{i}^{2}\varepsilon
\left( \omega _{i},\mathbf{r}\right) \right] }{\omega _{i}\partial \omega
_{i}}\mathbf{E}_{i}^{\ast }\left( \mathbf{r,}p\right) \mathbf{E}_{i}\left(
\mathbf{r,}p\right) d^{3}r=4\pi \hbar \omega _i (p).
\label{NC}
\end{equation}%
The solution of the Schr\"{o}dinger
equation $i\hbar \frac{\partial }{\partial t}\left\vert \Psi
_{i}\right\rangle =\hat{H}_{i}\left\vert \Psi _{i}\right\rangle $ for a
given field mode is
\begin{equation}
\left\vert \Psi _{i}\right\rangle =\sum_{n=0}^{\infty
}   C_{n}\left\vert n\right\rangle  \label{Sol of SE}
\end{equation}%
where $C_n = C_n^0 e^{-i(n + \frac{1}{2}) \int_{0}^{t}\omega _{i}\left( \tau \right) d\tau
}$, $\omega _{i}\left( t\right) \equiv $ $\omega _{i%
}\left( p\left( t\right) \right) $, and $\left\vert n\right\rangle $ are Fock
states. For a bosonic field described by a standard quantized harmonic oscillator, if we choose the coordinate representation expressed via
Hermite polynomials, the parameters of the polynomials will be
time-dependent. One can easily see that the above solution conserves the
adiabatic invariant $\frac{\left\langle \Psi _{i}\right\vert \hat{H}%
_{i}\left\vert \Psi _{i}\right\rangle }{\omega _{i}\left( t\right) }$
, just like in a classical slowly time-varying harmonic oscillator problem \cite{LL1}.

\subsection{Quantum emitter coupled to a quantized EM field
with a time-variable amplitude}

Let a two-level electron system (an atom)  be located at the point $\mathbf{r}=\mathbf{0}$
inside the cavity. The Hamiltonian of the system including the coupling of
an atom to a particular cavity mode and its adiabatic modulation can be
described within the RWA,
\begin{equation}
\hat{H}=\hbar \omega(t) \left( \hat{c}^{\dagger }%
\hat{c}+\frac{1}{2}\right) +W\hat{\sigma}^{\dagger }\hat{\sigma}-\left[
\mathbf{d}\cdot \mathbf{E}\left( t\right) \hat{\sigma}^{\dagger }\hat{c}+%
\mathbf{d}^{\ast }\cdot \mathbf{E}^{\ast }\left( t\right) \hat{\sigma}\hat{c}%
^{\dagger }\right] ,  \label{H of sys with RWA}
\end{equation}%
where $\mathbf{E}\left( \mathbf{0},t\right) =\mathbf{E}\left( t\right) $. The time dependence of the field
amplitude follows from the parameter modulation.

The wave function of the coupled photon-electron state can be written as
\begin{equation}
\label{wavefun}
\Psi =\sum_{n=0}^{\infty } \left( C_{n0}\left\vert n\right\rangle \left\vert
0\right\rangle +C_{n1}\left\vert n\right\rangle \left\vert 1\right\rangle \right), 
\end{equation}
 where we will maintain the same order, $\left\vert
photon\right\rangle \left\vert fermion\right\rangle $ of the state products
everywhere. Substituting it in the Schr\"{o}dinger equation and taking into
account the time variation of the parameter, we obtain the equation for the ground energy state, 
\begin{equation}
\dot{C}_{00}+i\omega _{00}\left( t\right) C_{00}=0,  \label{Eq for les}
\end{equation}
and a pair of equations for ``resonant'' states,
\begin{equation}
\dot{C}_{n0}+i\omega _{n0}\left( t\right) C_{n0}-i\Omega _{R}^{\ast }\left(
t\right) C_{\left( n-1\right) 1}=0,  \label{Cn0 dot}
\end{equation}%
\begin{equation}
\dot{C}_{\left( n-1\right) 1}+i\omega _{\left( n-1\right) 1}\left( t\right)
C_{\left( n-1\right) 1}-i\Omega _{R}\left( t\right) C_{n0}=0,
\label{Cn-11 dot}
\end{equation}%
where
\begin{equation*}
\omega _{n0}\left( t\right) =\omega \left( t\right) \left( n+\frac{1}{2}%
\right) ,\ \ \omega _{n1}\left( t\right) =\omega _{n0}\left( t\right) +\frac{%
W}{\hbar },\ \ \Omega _{R}\left( t\right) =\frac{\mathbf{d\cdot E}\left(
t\right) }{\hbar }\sqrt{n}.
\end{equation*}%

Equations (\ref{Cn0 dot}), (\ref{Cn-11 dot}) can be written in a more convenient form after making a substitution
\begin{equation}
\left(
\begin{array}{c}
C_{n0} \\
C_{\left( n-1\right) 1}%
\end{array}%
\right) =\left(
\begin{array}{c}
G_{n0}e^{-i\int_{0}^{t}\omega _{n0}\left( \tau \right) d\tau } \\
G_{\left( n-1\right) 1}e^{-i\int_{0}^{t}\omega _{\left( n-1\right) 1%
}\left( \tau \right) d\tau }%
\end{array}%
\right) ,  \label{Cn0 and Cn-11}
\end{equation}%
which gives 
\begin{equation}
\dot{G}_{n0}-i\Omega _{R}^{\ast }\left( t\right) e^{i\int_{0}^{t}\delta
\left( \tau \right) d\tau }G_{\left( n-1\right) 1}=0,  
\label{Gn0 dot}
\end{equation}
\begin{equation}
\dot{G}_{\left( n-1\right) 1}-i\Omega _{R}\left( t\right)
e^{-i\int_{0}^{t}\delta \left( \tau \right) d\tau }G_{n0}=0,
\label{Gn-11 dot}
\end{equation}%
where $\delta \left( t\right) =\omega \left( t\right) -\frac{W}{\hbar }$.

When there is no modulation, i.e. $\delta$, $\omega$, and $\Omega_R$ are constant, Eqs.~(\ref{Gn0 dot}), (\ref{Gn-11 dot}) have a simple solution $G_{(n-1)1},G_{n0} \propto e^{-i\nu t}$, where the eigenvalues are 
\begin{equation}
    \nu_{1,2} = \frac{\delta}{2} \pm \sqrt{\frac{\delta^2}{4} + |\Omega_R|^2 }, 
    \label{nu12}
\end{equation}
and the eigenvectors satisfy 
\begin{equation}
    K_{1,2} = \left[ \frac{G_{n0}}{G_{(n-1)1}} \right]_{1,2} = \frac{\nu_{1,2} e^{i \delta t}}{\Omega_R},
\end{equation}
where $K_1 K_2^{\ast} = -1$. The eigenvalues $\nu_{1,2}$ as a function of detuning $\delta$ are shown in Fig.~2. It is easy to verify that in the region $\delta \ll -|\Omega_R|$   the eigenvalue $\nu_1$ corresponds to the dominant state $\left\vert n-1\right\rangle \left\vert 1\right\rangle$, whereas in the region $\delta \gg |\Omega_R|$   this eigenvalue corresponds to dominant state $\left\vert n\right\rangle \left\vert 0\right\rangle$. For the eigenvalue $\nu_2$ it is exactly the opposite.

 %%%%%%%%%%%%%%%%%%%%%%%%%%%%%%%

\begin{figure}[htb]
\centering
\begin{subfigure}[b]{0.8\textwidth}
\includegraphics[width=1\linewidth]{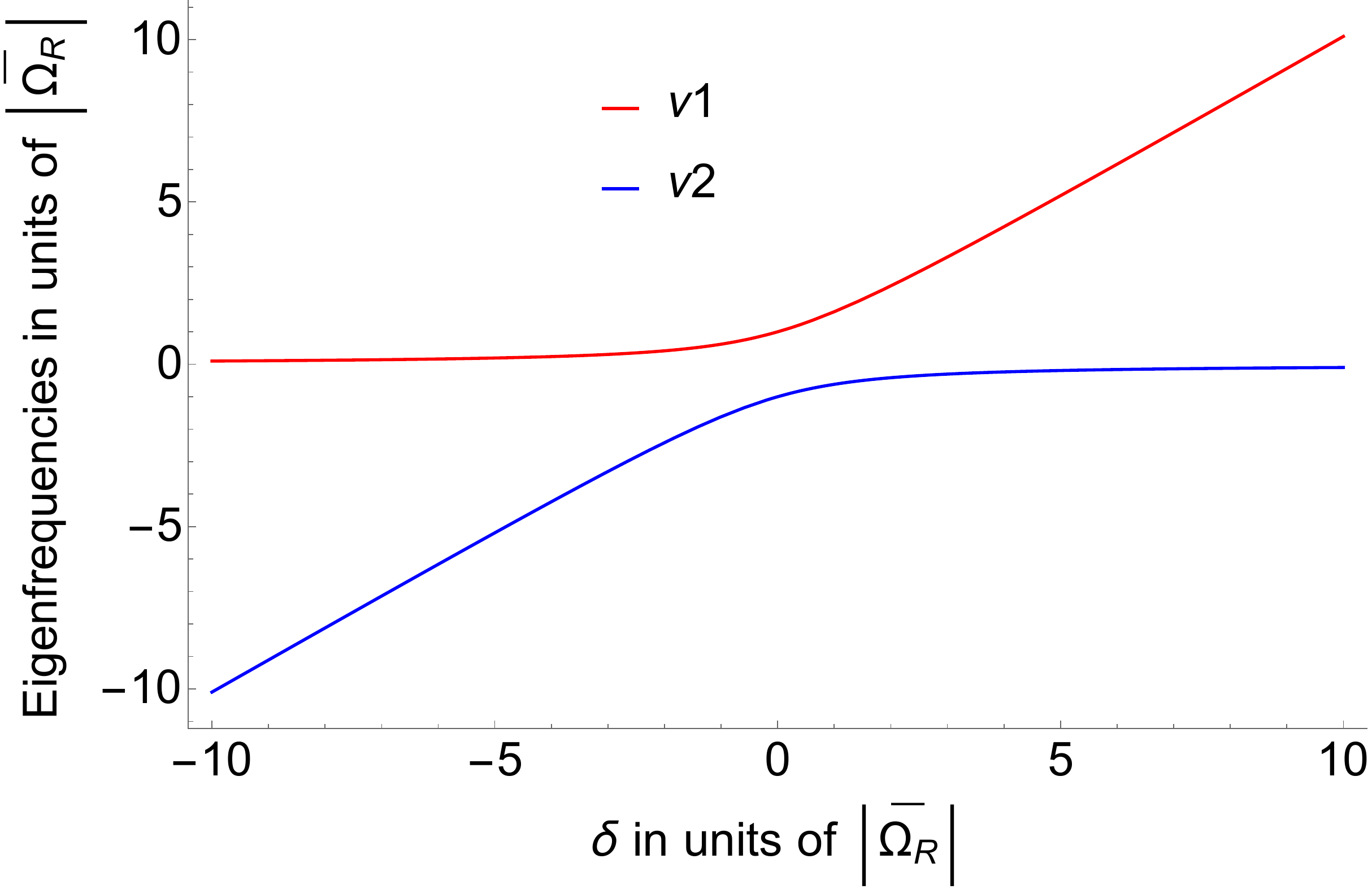}
\caption{ }
\label{fig2}
\end{subfigure}

\caption { Frequency eigenvalues $\nu_{1,2}$ from Eq.~(\ref{nu12}) as a function of detuning $\delta$ from the  resonance, $\delta = \omega - \frac{W}{\hbar}$. All frequencies are in units of the average Rabi frequency $\overline{\Omega _{R}}$.}
\end{figure}

%%%%%%%%%%%%%%%%%%%%%%%%%

When a cavity parameter is modulated, for example, a cavity height $d(t)$ in Fig.~1, both frequencies and field amplitudes $\mathbf{E}_{i}\left(\mathbf{r,}p\right)$ get modulated; see Eq.~(\ref{NC}). Therefore, the Rabi frequency $\Omega_R$ gets modulated. For a periodic modulation, the function $ \Omega _{R}\left( t\right)
e^{-i\int_{0}^{t}\delta \left( \tau \right) d\tau }$ is periodic and can be expanded in the Fourier series, 
\begin{equation}
    \Omega _{R}\left( t\right)
e^{-i\int_{0}^{t}\delta \left( \tau \right) d\tau } = \sum_{n = -\infty}^{\infty} R_n e^{-i n \Omega t},
\label{fourier}
\end{equation}
where $\Omega$ is the modulation frequency. The explicit expressions for the Fourier amplitudes $R_n$ can be obtained  for any specific model of a cavity; see, e.g., Appendix A for the plasmonic cavity, which shows specific examples of the cavity mode frequencies, field amplitudes, and their modulation.  

When the modulation frequency and amplitude of the eigenmode frequencies are small enough, one can neglect the modulation of the Rabi frequency in Eqs.~(\ref{Gn0 dot}), (\ref{Gn-11 dot}). This corresponds to the WKB approximation and one can see it by 
 taking the time derivative of Eq.~(\ref{Gn-11 dot}): 
\begin{equation}
\frac{d^{2}G_{\left( n-1\right) 1}}{dt^{2}} + \left( i\delta
\left( t\right) - \frac{1}{\Omega _{R}\left(
t\right) }  \frac{d\Omega _{R}}{dt} \right) \frac{dG_{\left( n-1\right) 1}}{dt} +\left\vert \Omega
_{R}\left( t\right) \right\vert ^{2}G_{\left( n-1\right) 1}=0.
\label{Gn-11 double dot}
\end{equation}
Now we can estimate the order of magnitude of different terms in Eq.~(\ref{Gn-11 double dot}).
Assume that the cavity mode frequency is modulated as $\omega_i(t) = \bar{\omega} + \delta(t) $. Since Eq.~(\ref{NC}) defines a certain dependence $\Omega_R(\omega_i)$, one can estimate $\left|\frac{\Omega_R^{-1} \dot{\Omega}_R}{\delta}\right| \sim \left| \frac{\dot{\delta}}{\Omega_R \delta} \frac{d \Omega_R}{d\omega_i}\right|_{\omega_i = \bar{\omega}}$ and $\left\vert \Omega _{R}-\overline{\Omega _{R}}\right\vert \sim \left| \frac{d\Omega_R}{d\omega_i} \delta \right|$. For estimations we take $\frac{d\Omega_R}{d\omega_i} \sim \frac{\overline{\Omega _{R}}}{\bar{\omega}} $, $\delta \sim \Delta \omega$, and $\dot{\delta} \sim \Omega \Delta \omega$ where $\overline{\Omega _{R}} = \Omega_R(\bar{\omega})$ and $\Delta \omega$ is the frequency change over the time $\Omega^{-1}$. This gives $\left|\frac{\Omega_R^{-1} \dot{\Omega}_R}{\delta}\right| \sim \frac{\Omega}{ \bar{\omega}}$ and $\left\vert \Omega _{R}-\overline{\Omega _{R}}\right\vert \sim \overline{%
\Omega _{R}}\frac{\Delta \omega }{\bar{\omega}}$. If $\Delta \omega ,\Omega
\ll \bar{\omega}$, Eq.~(\ref{Gn-11 double dot}) becomes
\begin{equation}
\frac{d^{2}G_{\left( n-1\right) 1}}{dt^{2}}+i\delta \left( t\right) \frac{%
dG_{\left( n-1\right) 1}}{dt}+\left\vert \overline{\Omega _{R}}\right\vert
^{2}G_{\left( n-1\right) 1}=0.  \label{Gn-11 double dot for sc}
\end{equation}%
Equation~(\ref{Gn-11 double dot for sc}) corresponds to the set of Eqs.~(\ref{Cn0 dot}),~(\ref{Cn-11 dot}) with $\Omega _{R} = $ const $ = \overline{\Omega _{R}}$.

If we consider for definiteness a sinusoidal
modulation of the frequency of a given mode, 
$ 
\omega \left( t\right) =\bar{\omega}-\Delta \omega \cdot \sin \left( \Omega
t\right)$, and take into account that $\Delta \omega, \Omega, \overline{\Omega_R} \ll \overline{\omega}$, the Fourier amplitudes in Eq.~(\ref{fourier}) can be expressed through the Bessel functions,  
\begin{equation}
    R_0 = \overline{\Omega _{R}} J_0\left(\frac{\Delta \omega}{\Omega}\right), \; R_n = (-i)^{|n|}  \overline{\Omega _{R}} J_{|n|}\left(\frac{\Delta \omega}{\Omega}\right).
    \label{amplitudes}
\end{equation}

The decoherence processes can be added within the stochastic equation of evolution for the state vector, which is derived in Appendix B. However, we postpone doing this until we consider a more complex case of two quantized modes interacting with a quantum emitter. 

\subsection{Simple manipulations with a qubit coupled to a single-mode field}

A single emitter coupled to a single-mode field in a time-variable cavity permits simple manipulations: a slow or fast sweep through the resonance $\omega(t) = W/\hbar$, bringing an electron-photon system in and out of entanglement by changing the values of coefficients in Eq.~(\ref{wavefun}), transduction of the excitation between an atom and the EM field, e.g., between  $\left\vert 0\right\rangle \left\vert
1 \right\rangle$ and $\left\vert 1\right\rangle \left\vert
0\right\rangle$ states, etc. 

Note that the rate of modulation or parameter variation has to be slow only as compared to the optical frequency. It does not have to be slow as compared to the average Rabi frequency $\overline{\Omega _{R}}$. Therefore, in the strong coupling regime a desired switching can be completed faster than the Rabi oscillations and decoherence rates. 

Let's look at some of these control operations in more detail. The sweep through resonance can be calculated exactly for each specific time dependence $\delta(t)$, but the limiting cases are well understood from the vast amount of literature on the linear coupling of the optical modes, Landau-Zener-type problems, etc \cite{kruskal1962,tokman2011,kochar1983,hallin1995,yokomizo2014}. 

For a slow sweep, $\left| \frac{d\delta}{dt} \right| \ll \overline{|\Omega_R|^2}$, the system will follow each eigenvalue branch plotted in Fig.~2 without jumping between them: for example, if the system starts from $\nu_1$ at $\delta \ll -\overline{|\Omega_R|}$, it will stay on $\nu_1$ as it moves through resonance to $\delta \gg \overline{|\Omega_R|}$. This means that the quantum state of the system will be switched from $\left\vert n-1\right\rangle \left\vert 1\right\rangle$ to $\left\vert n\right\rangle \left\vert 0\right\rangle$. 

In the opposite limit of a fast sweep, $\left| \frac{d\delta}{dt} \right| \gg \overline{|\Omega_R|^2}$, as the system moves through resonance from $\delta \ll -\overline{|\Omega_R|}$ to $\delta \gg \overline{|\Omega_R|}$ it jumps from one eigenvalue branch to another. As a result, the quantum state stays unchanged. 

In the intermediate region $\left| \frac{d\delta}{dt} \right| \sim \overline{|\Omega_R|^2}$, by varying the sweep rate or the Rabi frequency  $\overline{|\Omega_R|}$ one can get any desired combination of the quantum states at the output. In particular, for linear variation of the detuning, $\delta(t) = \beta t$ where $\beta$ is a constant, one can obtain an exact analytic solution of Eq.~(\ref{Gn-11 double dot for sc}) to predict the evolution of the system: 
\begin{equation}
G_{(n-1)1}(t) =\displaystyle  e^{-\frac{i \beta t^2}{4}} \left[ c_1 D_{i \frac{\overline{|\Omega_R|}^2}{\beta}} \left( \sqrt{\beta} e^{-\frac{i \pi}{4}} t \right) + c_2 D_{-i \frac{\overline{|\Omega_R|}^2}{\beta}-1} \left( i \sqrt{\beta} e^{-\frac{i \pi}{4}} t \right)  \right],
\label{linear}
\end{equation}
where $D_{\nu}$ are the parabolic cylinder functions \cite{bateman} and $c_{1,2}$ are arbitrary constants determined by initial conditions. This solution can be used, for example, to calculate the efficiency of the $\left\vert n-1\right\rangle \left\vert 1\right\rangle$ quantum state tunneling, i.e., the probability of the transition from the top to bottom branch in Fig.~2 as the detuning $\delta(t)$ varies from $-\infty$ to $+\infty$: 
$$
\left| C_{(n-1)1} \right|_{\delta \rightarrow \infty}^2 \approx \displaystyle e^{-\frac{\overline{2 \pi |\Omega_R|}^2}{\beta}} \left| C_{(n-1)1} \right|_{\delta \rightarrow - \infty}^2.
$$
As expected, the probability is approaching 1 when $\left| \frac{d\delta}{dt} \right| = \beta \gg \overline{|\Omega_R|}^2$ and becomes exponentially small in the opposite limit. 

%%%%%%%%%%%%%%%%%%%%%%%%%

\section{Dynamics of two modulated cavity modes coupled to a
quantum emitter}

In order to perform more complex operations on the photonic qubits and get more functionality, we need to add one more quantized degree of freedom to the system. Here we consider {\it two} cavity modes in a time-variable cavity,
\begin{equation}
\mathbf{\hat{E}}=\mathbf{E}_{a}\left( \mathbf{r,}t\right) \hat{a}+\mathbf{E}%
_{a}^{\ast }\left( \mathbf{r,}t\right) \hat{a}^{\dagger }+\mathbf{E}%
_{b}\left( \mathbf{r,}t\right) \hat{b}+\mathbf{E}_{b}^{\ast }\left( \mathbf{%
r,}t\right) \hat{b}^{\dagger }.  \label{two mode ef}
\end{equation}%

We assume that the modulation of both frequencies has a small amplitude and average frequencies of both modes $\bar{\omega}_{a,b}$
are close to the transition frequency. In this case the RWA Hamiltonian for
an atom + field system is
\begin{equation}
\hat{H}=\hbar \omega _{a}\left( t\right) \left( \hat{a}^{\dagger }%
\hat{a}+\frac{1}{2}\right) +\hbar \omega _{b}\left( t\right) \left(
\hat{b}^{\dagger }\hat{b}+\frac{1}{2}\right) +W\hat{\sigma}^{\dagger }\hat{%
\sigma}-\left[ \hat{\sigma}^{\dagger }\left( \chi _{a}\hat{a}+\chi _{b}\hat{b%
}\right) +\hat{\sigma}\left( \chi _{a}^{\ast }\hat{a}^{\dagger }+\chi
_{b}^{\ast }\hat{b}^{\dagger }\right) \right] ,  \label{toh in RWA}
\end{equation}%
where $\chi _{a,b}\left( t\right) =\mathbf{d\cdot E}_{a,b}\left( t\right) $.

The Schr\"{o}dinger equation can be solved analytically within the RWA
\cite{tokman2020}. As a simple example, we include only the transitions between
the states with lowest energies, namely $\left\vert 0_{a}\right\rangle
\left\vert 0_{b}\right\rangle \left\vert 0\right\rangle ,\left\vert
0_{a}\right\rangle \left\vert 0_{b}\right\rangle \left\vert 1\right\rangle
,\left\vert 1_{a}\right\rangle \left\vert 0_{b}\right\rangle \left\vert
0\right\rangle ,\left\vert 0_{a}\right\rangle \left\vert 1_{b}\right\rangle
\left\vert 0\right\rangle $, i.e. we seek the solution in the form%
\begin{equation}
\Psi =C_{000}\left\vert 0_{a}\right\rangle \left\vert 0_{b}\right\rangle
\left\vert 0\right\rangle +C_{001}\left\vert 0_{a}\right\rangle \left\vert
0_{b}\right\rangle \left\vert 1\right\rangle +C_{100}\left\vert
1_{a}\right\rangle \left\vert 0_{b}\right\rangle \left\vert 0\right\rangle
+C_{010}\left\vert 0_{a}\right\rangle \left\vert 1_{b}\right\rangle
\left\vert 0\right\rangle.   \label{sol of SE}
\end{equation}

For arbitrary coefficients $C$ the state (\ref{sol of SE}) is a tripartite entangled state which can be reduced to standard GHZ states by  local operations  \cite{dur2000, cunha2020}, e.g. by rotations on the Bloch sphere of each qubit. In most cases discussed in the literature the GHZ states are made of identical subsystems, e.g., photons \cite{shalm2012, agusti2020}. In our case the subsystems are of different nature: a fermionic electron system and bosonic EM field modes. This makes their rotations more complicated, but on the other hand, enables other interesting applications. For example, one can determine the statistics of atomic excitations by measuring the statistics of photons, or change the entangled state of coupled photon modes by changing the atomic state with a classical control field.  

Similarly to \cite{tokman2020}, the equations for the coefficients are 
\begin{equation}
\dot{C}_{000}+i\frac{\omega _{a}\left( t\right) +\omega _{b}\left( t\right)
}{2}C_{000}=0;  \label{c000 dot}
\end{equation}%
\begin{equation}
\dot{C}_{001}+i\left( \frac{1}{2}\omega _{a}\left( t\right) +\frac{1}{2}%
\omega _{b}\left( t\right) +\frac{W}{\hbar }\right) C_{001}-i\Omega
_{Ra}(t) C_{100}-i\Omega _{Rb}(t) C_{010}=0,  \label{c001 dot}
\end{equation}%
\begin{equation}
\dot{C}_{100}+i\left( \frac{3}{2}\omega _{a}\left( t\right) +\frac{1}{2}%
\omega _{b}\left( t\right) \right) C_{100}-i\Omega _{Ra}^{\ast }(t)C_{001}=0,
\label{c100 dot}
\end{equation}%
\begin{equation}
\dot{C}_{010}+i\left( \frac{1}{2}\omega _{a}\left( t\right) +\frac{3}{2}%
\omega _{b}\left( t\right) \right) C_{010}-i\Omega _{Rb}^{\ast }(t)C_{001}=0,
\label{c010 dot}
\end{equation}%
where $\Omega _{Ra,b}=\frac{\chi _{a,b}}{\hbar }$. Making the substitution%
\begin{equation}
\left(
\begin{array}{c}
C_{001} \\
C_{100} \\
C_{010}%
\end{array}%
\right) =\left(
\begin{array}{c}
G_{0}\exp \left[ -i\int_{0}^{t}\left( \frac{1}{2}\omega _{a}\left( \tau
\right) +\frac{1}{2}\omega _{b}\left( \tau \right) +\frac{W}{\hbar }\right)
d\tau \right] \\
G_{a}\exp \left[ -i\int_{0}^{t}\left( \frac{3}{2}\omega _{a}\left( \tau
\right) +\frac{1}{2}\omega _{b}\left( \tau \right) \right) d\tau \right] \\
G_{b}\exp \left[ -i\int_{0}^{t}\left( \frac{1}{2}\omega _{a}\left( \tau
\right) +\frac{3}{2}\omega _{b}\left( \tau \right) \right) d\tau \right]%
\end{array}%
\right) ,  \label{substitution}
\end{equation}%
we obtain%
\begin{equation}
\dot{G}_{0}-i\Omega _{Ra}(t)G_{a}\exp \left[ -i\int_{0}^{t}\left( \omega
_{a}\left( \tau \right) -\frac{W}{\hbar }\right) d\tau \right] -i\Omega
_{Rb}(t) G_{b}\exp \left[ -i\int_{0}^{t}\left( \omega _{b}\left( \tau \right) -%
\frac{W}{\hbar }\right) d\tau \right] =0,  \label{g0 dot}
\end{equation}%
\begin{equation}
\dot{G}_{a}-i\Omega _{Ra}^{\ast }(t) G_{0}\exp \left[ i\int_{0}^{t}\left( \omega
_{a}\left( \tau \right) -\frac{W}{\hbar }\right) d\tau \right] =0,
\label{ga dot}
\end{equation}%
\begin{equation}
\dot{G}_{b}-i\Omega _{Rb}^{\ast }(t) G_{0}\exp \left[ i\int_{0}^{t}\left( \omega
_{b}\left( \tau \right) -\frac{W}{\hbar }\right) d\tau \right] =0,
\label{gb dot}
\end{equation}%

\begin{figure}[htb]
	\centering
	\begin{subfigure}{0.45\textwidth}
		\centering
		\includegraphics[width=\linewidth]{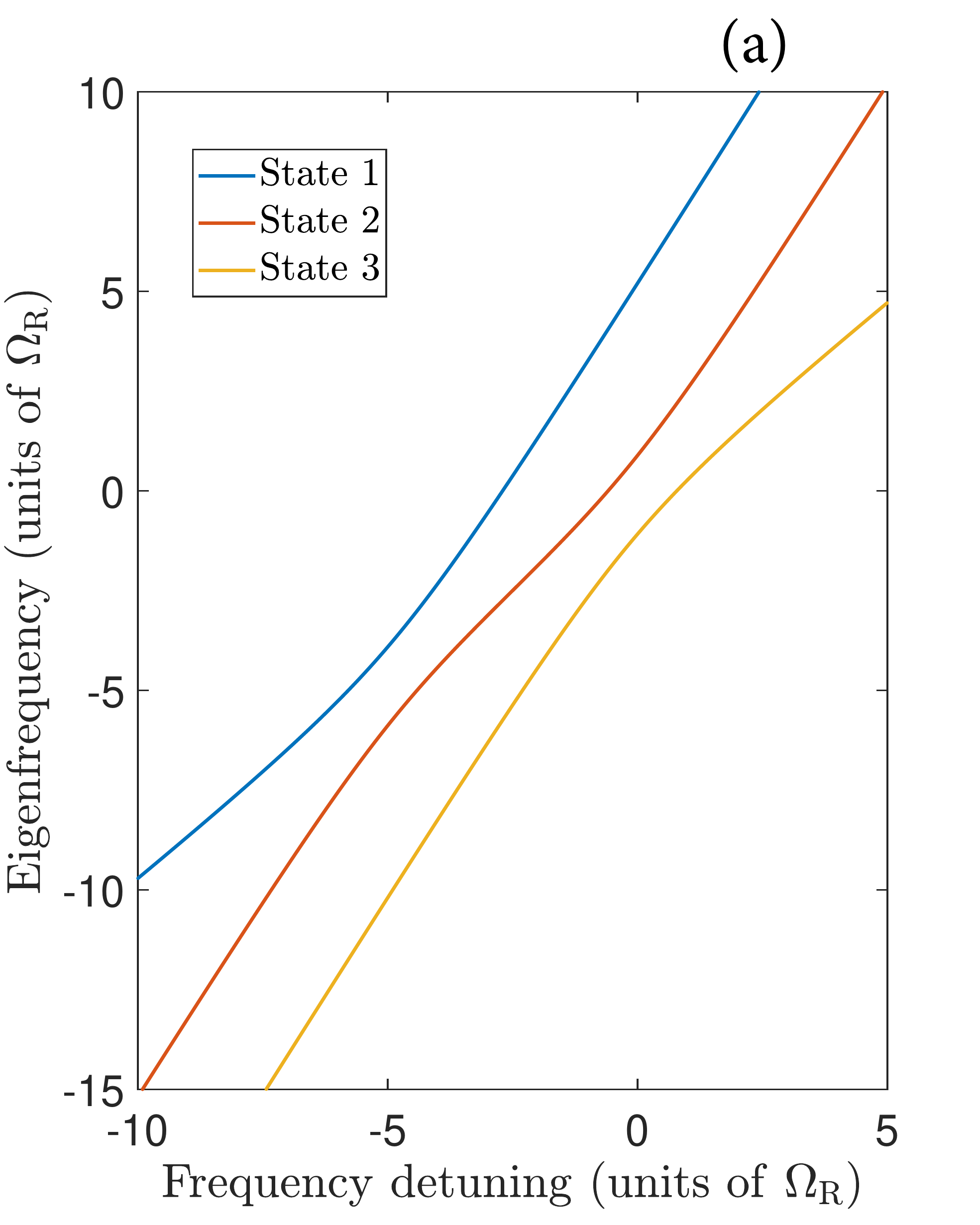}
	\end{subfigure}
	\begin{subfigure}{0.45\textwidth}
		\includegraphics[width=\linewidth]{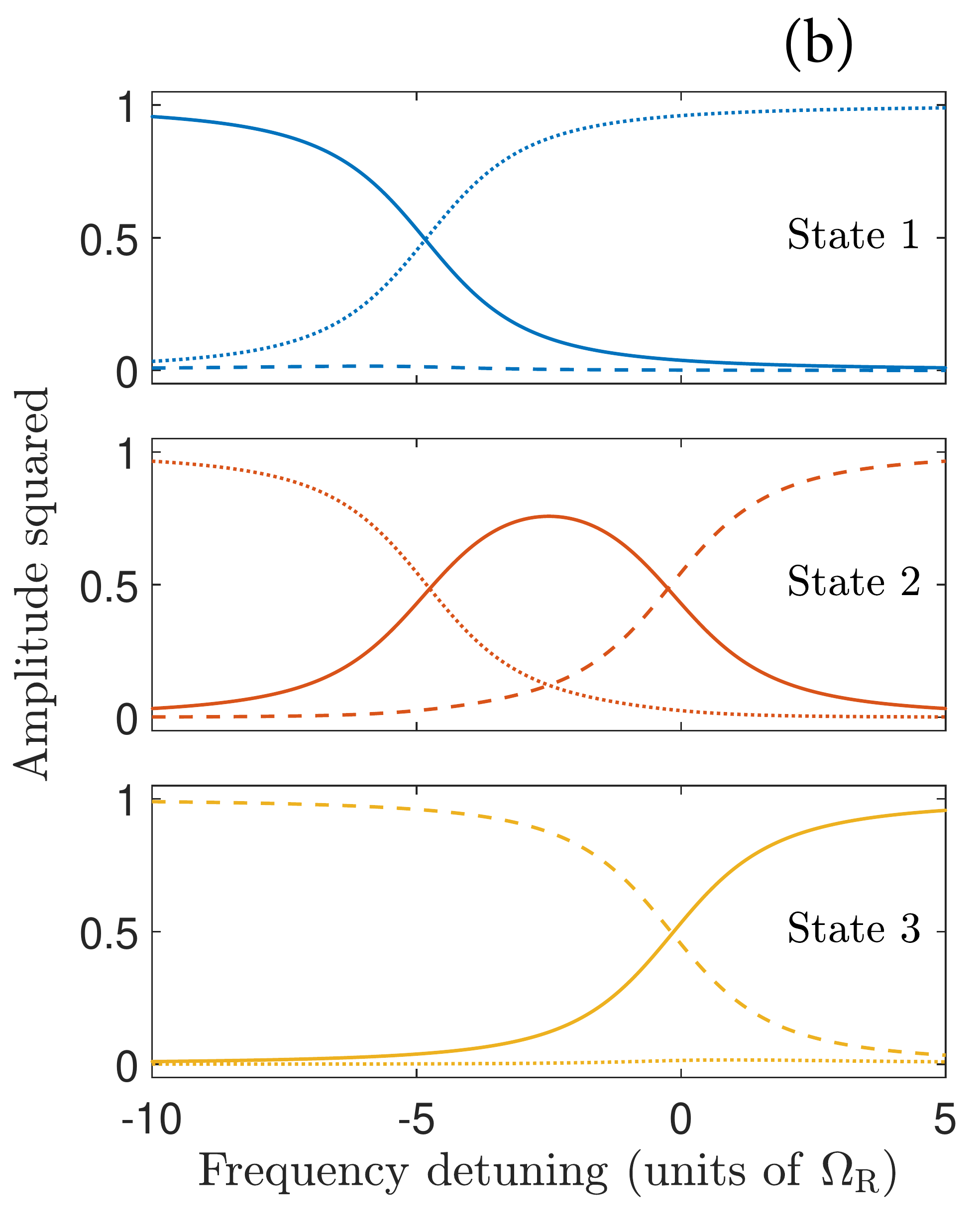}
	\end{subfigure}
	\caption{ The eigenstates of the system described by Eqs.~\eqref{c001 dot}, \eqref{c100 dot} and \eqref{c010 dot} as a function of frequency detuning defined as $\omega_a-W/\hbar$, whereas the difference of modal frequencies $\omega_b - \omega_a = 5\Omega_{R}$ is kept constant. The eigenfrequencies are shown in (a), and the amplitudes of the eigenstates are shown in (b), in which the amplitudes of $C_{001}$, $C_{100}$ and $C_{010}$ are represented by the solid, dashed and dotted lines, respectively. The eigenfrequencies are shifted by $\left.\left( \frac{1}{2}\omega_a + \frac{1}{2}\omega_b + \frac{W}{\hbar} \right)\right|_{\omega_a=W/\hbar}$. } 
	\label{Fig:eigen_2photon}
\end{figure}

In Fig.~\ref{Fig:eigen_2photon}, we show the eigenstates of the system described by Eqs.~\eqref{c001 dot}, \eqref{c100 dot} and \eqref{c010 dot} as a function of frequency detuning defined as $\omega_a-W/\hbar$.  Here we assumed that $\Omega_{Ra}=\Omega_{Rb}\equiv \Omega_R$ and kept the difference  $\omega_b - \omega_a = 5\Omega_{R}$ constant, which can be achieved either by varying $W/\hbar$ while keeping constant $\omega_{a,b}$ or by varying $\omega_a$ and $\omega_b$ at the same rate while keeping $W/\hbar$ constant. The anticrossings are clearly seen in the plot of eigenfrequencies, when either $\omega_a$ or $\omega_b$ is resonant with the optical transition of an atom. As compared to Fig.~2, Fig.~\ref{Fig:eigen_2photon}(b) shows more possibilities for switching between the three product states as the detuning is swept through the two resonances at the rate slower than the Rabi frequencies and the generation of both bipartite and tripartite entangled states in the vicinity of resonances if the sweeping rate is comparable to the Rabi frequencies. 

%%%%%%%%%%%%%%%%%%%%%%%%%%%%%%%%%%%%

Since the functions $\omega_{a,b}(t)$ and $\Omega_{a,b}$ are periodic with period $2 \pi/\Omega$, we can use the expansion (\ref{fourier}) in Eqs.~(\ref{g0 dot})-(\ref{gb dot}).

If we keep only the resonant terms, assuming for example the following
resonances, $\bar{\omega}_{a}=\frac{W}{\hbar }$ and $\bar{\omega}%
_{b}+m\Omega =\frac{W}{\hbar }$, where $m$ is the number of a particular Fourier harmonic, the equations get simplified, 
\begin{equation}
\frac{d}{dt}\left(
\begin{array}{c}
G_{0} \\
G_{a} \\
G_{b}%
\end{array}%
\right) +\left(
\begin{array}{ccc}
0 & -i R_{a0} &
-i R_{bm} \\
-iR_{a0}^{\ast } & 0 & 0 \\
-i R_{bm}^{\ast} & 0 & 0
\end{array}%
\right) \left(
\begin{array}{c}
G_{0} \\
G_{a} \\
G_{b}%
\end{array}%
\right) =0.  \label{simp eq}
\end{equation}%

Other (nonresonant) harmonics can be neglected only if $\Omega_{Ra,b} \ll \Omega$, see \cite{tokman2020}. When the modulation amplitude is zero, $R_{a0} = \Omega_{Ra}$ and $R_{bm} = 0$. In this case one of the eigenvalues $\Gamma_0$ corresponds to the decoupled state  $\left\vert 0_{a}\right\rangle \left\vert 1_{b}\right\rangle \left\vert
0\right\rangle$. Two other eigenvalues $\Gamma_{1,2}$ describe the solution with Rabi oscillations between states $\left\vert 0_{a}\right\rangle \left\vert 0_{b}\right\rangle \left\vert
1\right\rangle$ and $\left\vert 1_{a}\right\rangle \left\vert 0_{b}\right\rangle \left\vert
0\right\rangle$.This is an obvious limit since frequency $\omega_a$ is in resonance with the transition frequency, whereas  $\omega_b$ is out of resonance. 

Assuming a sinusoidal
modulation of the partial frequencies of both cavity modes as an example,
\begin{equation}
\omega _{a,b}\left( t\right) =\bar{\omega}_{a,b}-\Delta \omega _{a,b}\cdot
\sin \left( \Omega t\right) ,  
\label{two mode sin modul of freq}
\end{equation}%
 and using the well-known expansion in series of the
harmonics of the modulation frequency $\Omega $, with coefficients expressed
in terms of Bessel functions,
\begin{equation}
e^{-i\frac{\Delta \omega }{\Omega }\cos \left( \Omega t\right) }=J_{0}\left(
\frac{\Delta \omega }{\Omega }\right) +2\sum_{n=1}^{\infty }\left( -i\right)
^{n}J_{n}\left( \frac{\Delta \omega }{\Omega }\right) \cos \left( n\Omega
t\right),  \label{coe}
\end{equation}
we can express Fourier amplitudes in Eq.~(\ref{simp eq}) through Bessel functions:
\begin{equation}
R_{a0} = \overline{\Omega _{Ra}} J_0\left(\frac{\Delta \omega_a}{\Omega}\right), \; R_{bm} = (-i)^{|m|}  \overline{\Omega _{Rb}} J_{|m|}\left(\frac{\Delta \omega_b}{\Omega}\right).
    \label{amplitudes2}
\end{equation}

Note that the modulation amplitudes in Eq.~(\ref{two mode sin modul of freq}) \textit{can be of
the order of the modulation frequency}, $\frac{\Delta \omega _{a,b}}{\Omega }%
\sim 1$, despite the requirement $\Delta \omega _{a,b}\ll $ $\bar{\omega}%
_{a,b}$.

As usual, to solve Eq.~(\ref{simp eq}) one has to find the
eigenvalues $\Gamma _{0,1,2}$ and eigenvectors of the matrix of
coefficients. The characteristic equation for the eigenvalues is $\Gamma
\left( \Gamma ^{2}+\Omega _{R\Sigma }^{2}\right) =0$, where the cumulative Rabi frequency is 
\begin{equation}
\label{cum-rabi} 
\Omega
_{R\Sigma }=\sqrt{|R_{a0}|^2 + |R_{bm}|^2 }.
\end{equation}
 The result is 
\begin{equation}
\Gamma _{0}=0,\ \Gamma _{1,2}=\pm i\Omega _{R\Sigma }.  \label{eigenvalues}
\end{equation}

Figure 4 shows one example of the cumulative Rabi frequency $\Omega
_{R\Sigma }$ as a function of $\Delta \omega  = \Delta \omega _{a} = \Delta \omega _{b}$ for $m = 1$ and $\Omega _{Ra} = \Omega _{Rb}$. As expected, $\Omega_{R\Sigma }$ decays with detuning from resonances but the decay is nonmonotonic and depends on the order of harmonic resonances.

 %%%%%%%%%%%%%%%%%%%%%%%%%%%%%%%

\begin{figure}[htb]
	\includegraphics[width=0.5\textwidth]{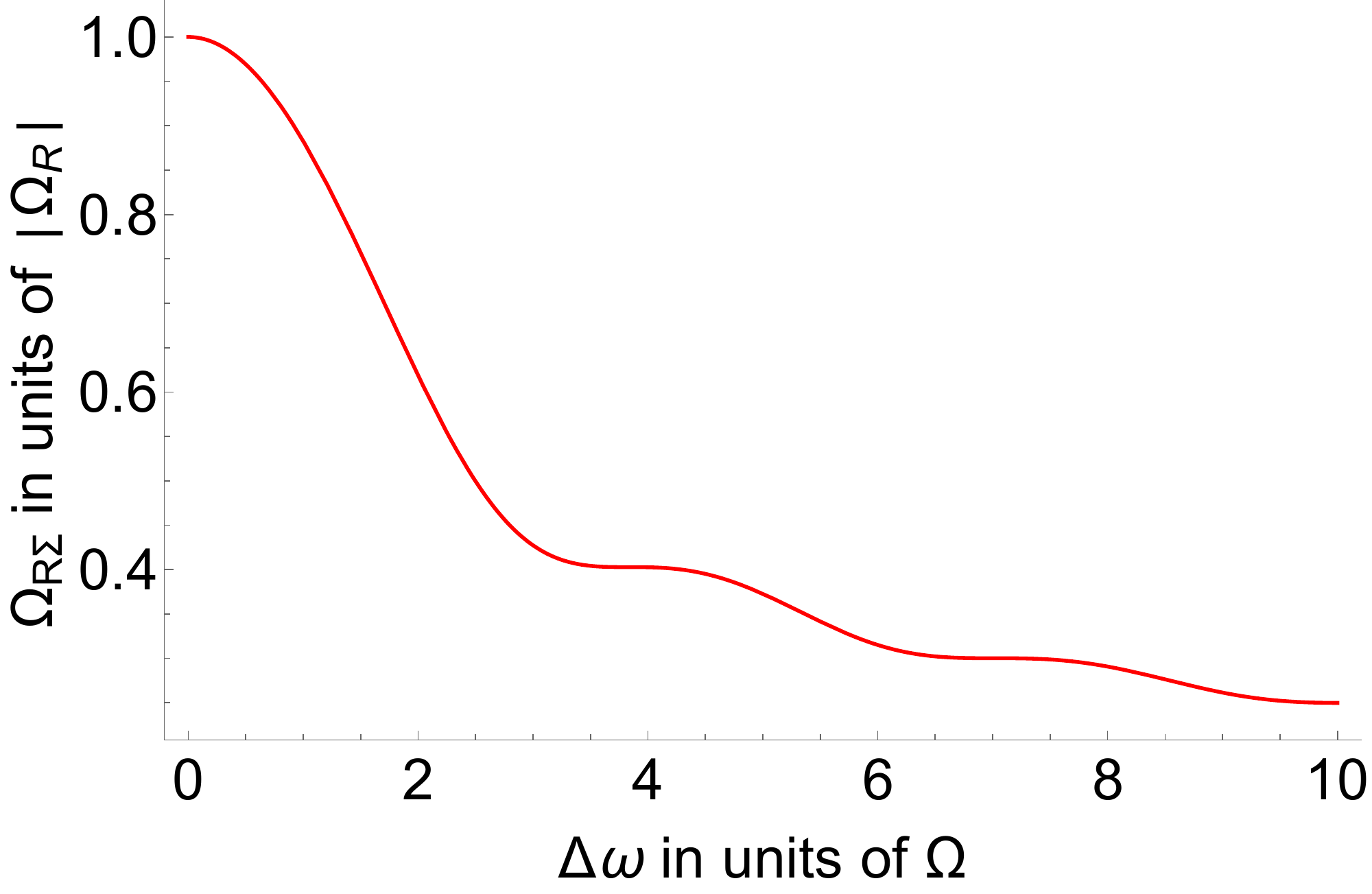}
\caption { Cumulative Rabi frequency $\Omega
_{R\Sigma }$ as a function of $\Delta \omega  = \Delta \omega _{a} = \Delta \omega _{b}$ for $m = 1$ and $\Omega _{R} = \Omega _{Ra} = \Omega _{Rb}$.}
\end{figure}

%%%%%%%%%%%%%%%%%%%%%%%%%

The eigenvalue $\Gamma _{0}$ (i.e. the solution $\propto e^{-\Gamma _{0}t}$
) corresponds to the eigenvector $\left(
\begin{array}{c}
0 \\
1 \\
- \frac{R_{a0}}{R_{bm}} 
\end{array}%
\right) $, whereas eigenvalues $\Gamma _{1,2}$ (i.e. the solution behaving
as $\propto e^{-\Gamma _{1,2}t}$) correspond to the eigenvectors $\left(
\begin{array}{c}
\frac{\pm \Omega _{R\Sigma }}{R_{a0}^{\ast }} \\
1 \\
\frac{R_{bm}^{\ast}}{R_{a0}^{\ast}}.
\end{array}%
\right) $. Here the eigenvectors are not normalized to $1$. The resulting
solution is
\begin{equation}
\left(
\begin{array}{c}
G_{0} \\
G_{a} \\
G_{b}%
\end{array}%
\right) =A\left(
\begin{array}{c}
0 \\
1 \\
- \frac{R_{a0}}{R_{bm}} %
\end{array}%
\right) +B e^{-i\Omega _{R\Sigma }t}  \left(
\begin{array}{c}
\frac{ \Omega _{R\Sigma }}{R_{a0}^{\ast }} \\
1 \\
\frac{R_{bm}^{\ast}}{R_{a0}^{\ast}}.
\end{array}%
\right)
+C e^{i\Omega _{R\Sigma }t} \left(
\begin{array}{c}
\frac{- \Omega _{R\Sigma }}{R_{a0}^{\ast }} \\
1 \\
\frac{R_{bm}^{\ast}}{R_{a0}^{\ast}}.
\end{array}%
\right) ,  \label{sol of simp eq}
\end{equation}%
where the constants $A$, $B$ and $C$ are determined by the initial
conditions. 

For an arbitrary initial state vector 
\begin{equation}
\Psi =C_{000}(0)\left\vert 0_{a}\right\rangle \left\vert 0_{b}\right\rangle
\left\vert 0\right\rangle +C_{001}(0)\left\vert 0_{a}\right\rangle \left\vert
0_{b}\right\rangle \left\vert 1\right\rangle +C_{100}(0)\left\vert
1_{a}\right\rangle \left\vert 0_{b}\right\rangle \left\vert 0\right\rangle
+C_{010}(0)\left\vert 0_{a}\right\rangle \left\vert 1_{b}\right\rangle
\left\vert 0\right\rangle,
   \label{WF-init}
\end{equation}
satisfying the normalization condition
$$
|C_{000}(0)|^2 + |C_{001}(0)|^2 + |C_{100}(0)|^2 + |C_{010}(0)|^2 = 1,
$$
the constants in Eq.~(\ref{sol of simp eq}) are
\begin{align}
    & A = \frac{C_{100}(0) \frac{|R_{bm}|^2}{|R_{a0}|^2} - C_{010}(0) \frac{R_{bm}}{R_{a0}} }{1 + \frac{|R_{bm}|^2}{|R_{a0}|^2}}, \nonumber \\
    & B = \frac{1}{2} \left( \frac{C_{100}(0) + C_{010}(0) \frac{R_{bm}}{R_{a0}} }{1 + \frac{|R_{bm}|^2}{|R_{a0}|^2}} + C_{001}(0) \frac{R_{a0}^{\ast}}{\Omega_{R\Sigma}} \right), \nonumber \\
    & C = \frac{1}{2} \left( \frac{C_{100}(0) + C_{010}(0) \frac{R_{bm}}{R_{a0}} }{1 + \frac{|R_{bm}|^2}{|R_{a0}|^2}} - C_{001}(0) \frac{R_{a0}^{\ast}}{\Omega_{R\Sigma}} \right). 
 \label{abc}   
    \end{align}

Let's consider some examples of the initial conditions to
illustrate this solution.

\subsection{ An atom is excited; both modes are in the vacuum state:}

The initial state vector is $\Psi(0)
= \left\vert 0_{a}\right\rangle \left\vert 0_{b}\right\rangle
\left\vert 1\right\rangle$. In this case Eq.~(\ref{abc}) gives $A=0,B=-C=\frac{R_{a0}^{\ast} }{2\Omega
_{R\Sigma }}$. The full expression for the state vector at any moment of time becomes 
\begin{eqnarray}
\Psi &=& e^{-i\int_{0}^{t}\omega _{001}\left( \tau \right) d\tau
}\cos \left( \Omega _{R\Sigma }t\right) \left\vert 0_{a}\right\rangle
\left\vert 0_{b}\right\rangle \left\vert 1\right\rangle -i\frac{R_{a0}^{\ast }}{\Omega
_{R\Sigma }} e^{-i\int_{0}^{t}\omega _{100}\left( \tau \right) d\tau
}\sin \left( \Omega _{R\Sigma }t\right) \left\vert 1_{a}\right\rangle
\left\vert 0_{b}\right\rangle \left\vert 0\right\rangle  \notag \\
&&- i \frac{R_{bm}^{\ast } }{\Omega _{R\Sigma }}e^{-i\int_{0}^{t}\omega _{010%
}\left( \tau \right) d\tau }\sin \left( \Omega _{R\Sigma }t\right)
\left\vert 0_{a}\right\rangle \left\vert 1_{b}\right\rangle \left\vert
0\right\rangle,  \label{sv for excited atom}
\end{eqnarray}
where 
\begin{equation*}
\omega _{001}\left( t\right) =\frac{1}{2}\omega _{a}\left( t\right) +\frac{1%
}{2}\omega _{b}\left( t\right) +\frac{W}{\hbar },\ \ \omega _{100}\left(
t\right) =\frac{3}{2}\omega _{a}\left( t\right) +\frac{1}{2}\omega
_{b}\left( t\right) ,\ \ \ \ \omega _{010}\left( t\right) =\frac{1}{2}\omega
_{a}\left( t\right) +\frac{3}{2}\omega _{b}\left( t\right) .
\end{equation*}%

As we see, an initial atomic excitation decays into a pair of
electromagnetic modes. Their frequencies are modulated due to the modulation
of the cavity geometry and are split by the cumulative Rabi frequency.  
In the absence of dissipation the excitation energy oscillates back and forth between an atom and the field modes at the cumulative Rabi frequency.

\subsection{Both cavity modes are excited; the atom is in the ground state:}  

The initial state vector is $\Psi(0) =C_{100}(0)\left\vert
1_{a}\right\rangle \left\vert 0_{b}\right\rangle \left\vert 0\right\rangle
+C_{010}(0) \left\vert 0_{a}\right\rangle \left\vert 1_{b}\right\rangle
\left\vert 0\right\rangle$.  
In this case the state vector is 
\begin{eqnarray}
\Psi &=& - 2i B \frac{\Omega_{R\Sigma}}{R_{a0}^{\ast}} e^{-i\int_{0}^{t}\omega _{001}\left( \tau \right) d\tau
}\sin \left( \Omega _{R\Sigma }t\right) \left\vert 0_{a}\right\rangle
\left\vert 0_{b}\right\rangle \left\vert 1\right\rangle + \left( A + 2B\cos\left( \Omega _{R\Sigma }t\right) \right) e^{-i\int_{0}^{t}\omega _{100}\left( \tau \right) d\tau
} \left\vert 1_{a}\right\rangle
\left\vert 0_{b}\right\rangle \left\vert 0\right\rangle  \notag \\
&& + \left( 2 B \frac{R_{bm}^{\ast}}{R_{a0}^{\ast}}\cos \left( \Omega _{R\Sigma }t\right) -A\frac{R_{a0}}{R_{bm}} \right)  e^{-i\int_{0}^{t}\omega _{010
}\left( \tau \right) d\tau }
\left\vert 0_{a}\right\rangle \left\vert 1_{b}\right\rangle \left\vert
0\right\rangle,  \label{sv-excited-modes}
\end{eqnarray}
where 
\begin{align}
    & A = \frac{C_{100}(0) \frac{|R_{bm}|^2}{|R_{a0}|^2} - C_{010}(0) \frac{R_{bm}}{R_{a0}} }{1 + \frac{|R_{bm}|^2}{|R_{a0}|^2}}, \;
    B = C = \frac{1}{2}  \frac{C_{100}(0) + C_{010}(0) \frac{R_{bm}}{R_{a0}} }{1 + \frac{|R_{bm}|^2}{|R_{a0}|^2}}. 
 \label{abc2}   
    \end{align}

%%%%%%%%%%%%%%%%%%%%%%%%%%%%%%%

\begin{figure}[htb]
\centering

\begin{subfigure}[a]{0.4\textwidth}
\includegraphics[width=1\linewidth]{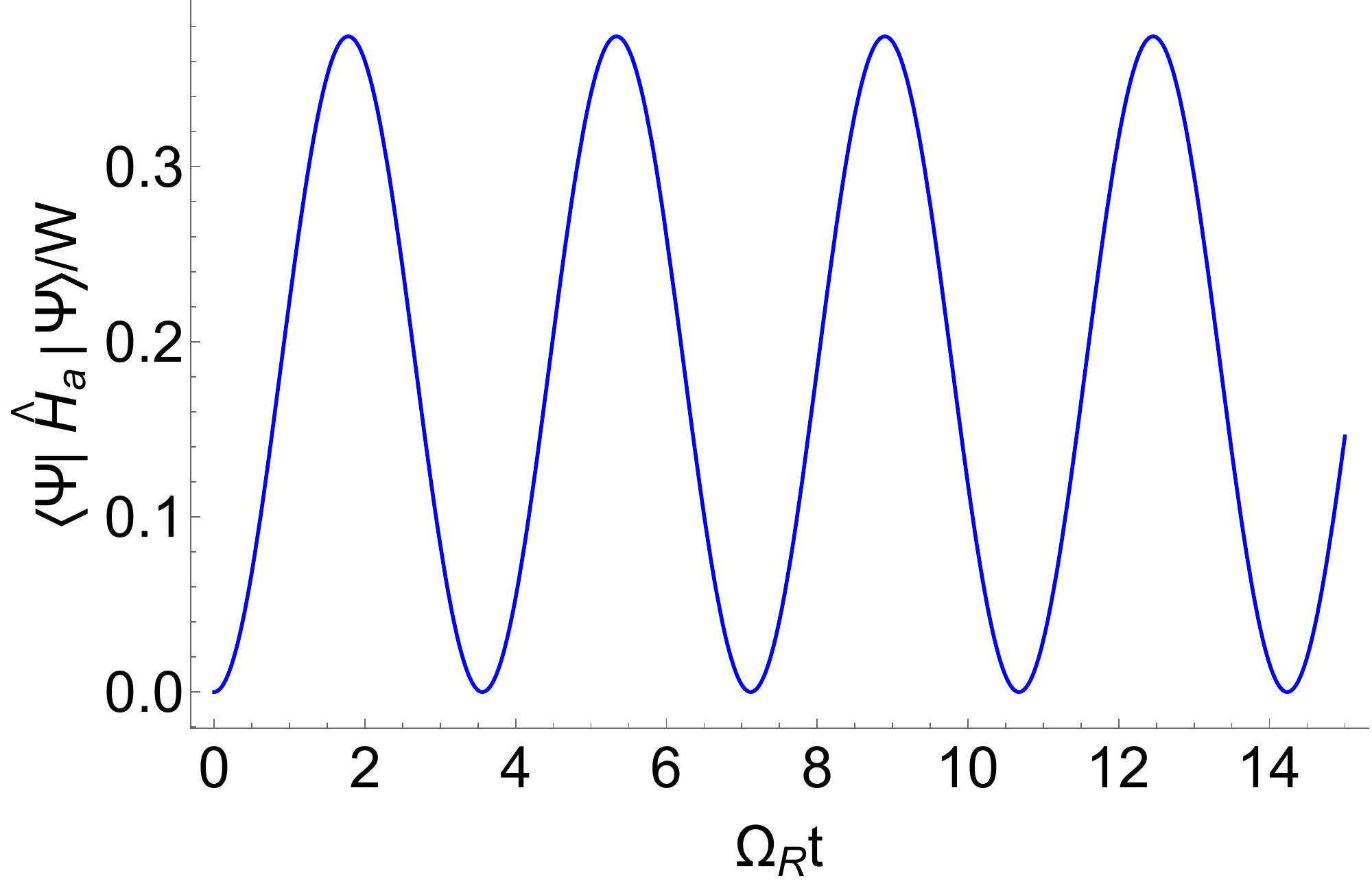}
\caption{ }
\end{subfigure}

\begin{subfigure}[b]{0.4\textwidth}
\includegraphics[width=1\linewidth]{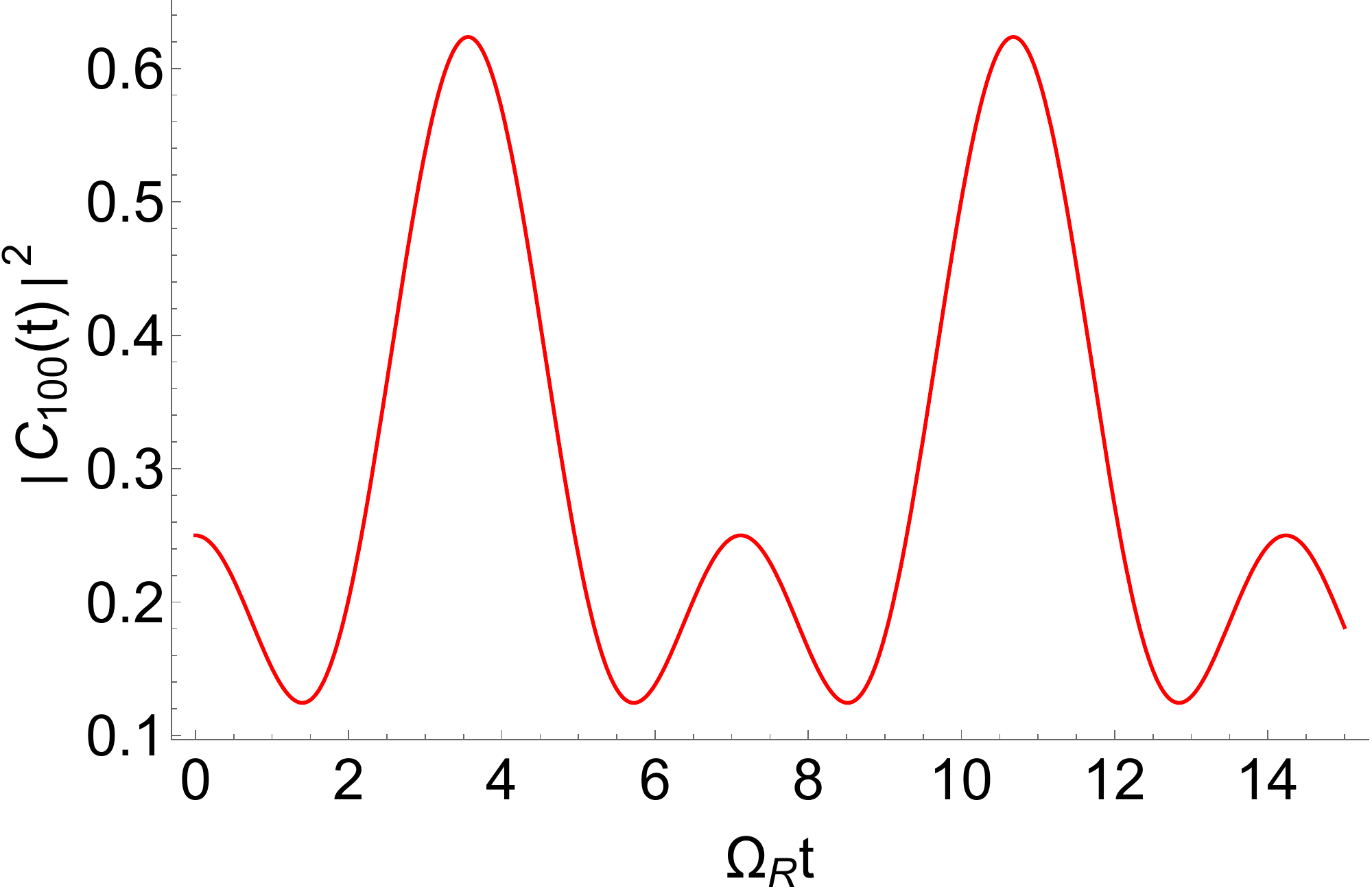}
\caption{}
\end{subfigure}

\begin{subfigure}[c]{0.4\textwidth}
\includegraphics[width=1\linewidth]{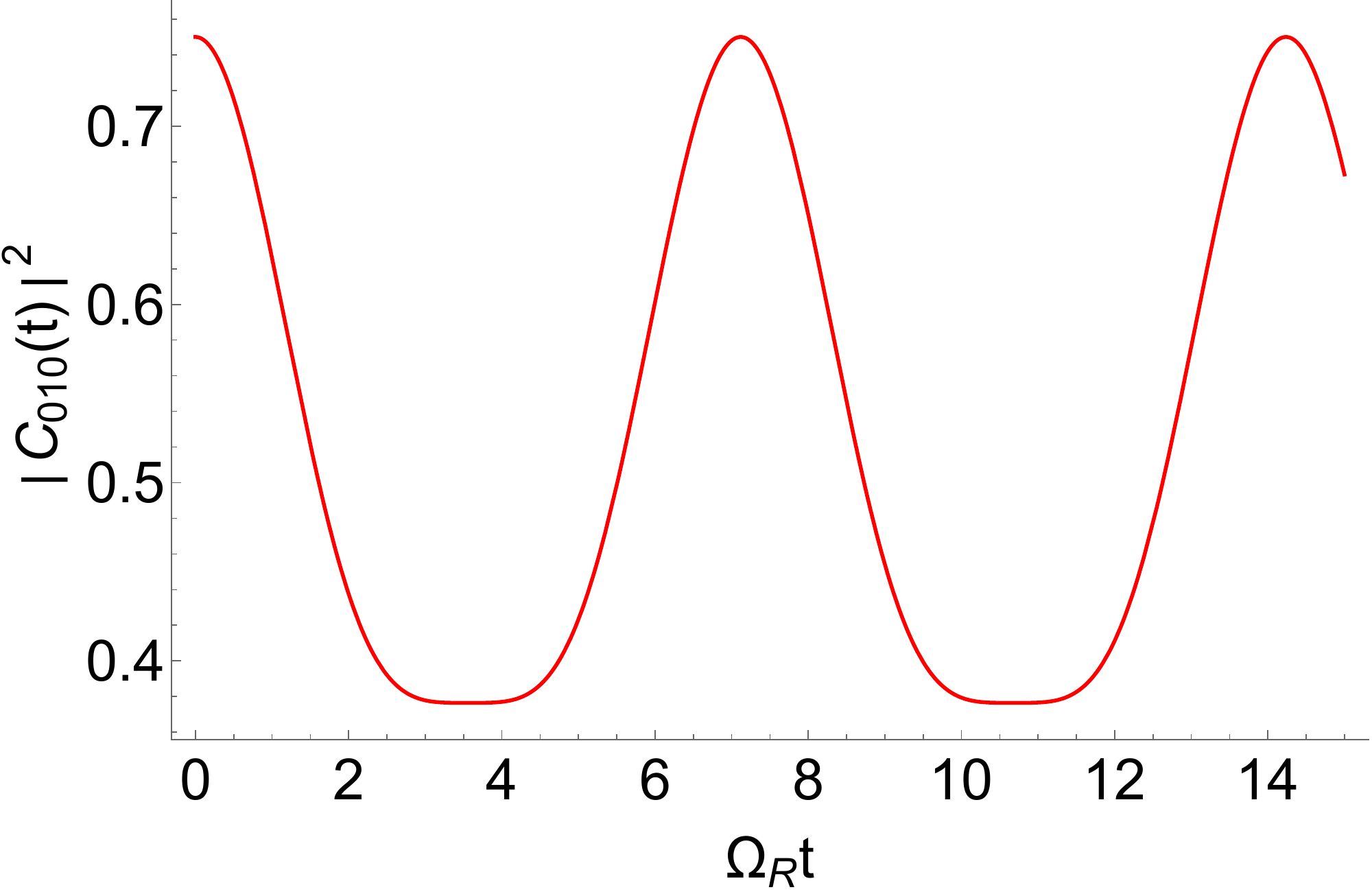}
\caption{}
\end{subfigure}

\caption {  (a) The average normalized energy of an atom, (b) the number of quanta in mode $a$, and (c) the number of quanta in mode $b$ as a function of normalized time. The initial conditions are $C_{000}(0) = 0$, $C_{001}(0) = 0$, $C_{100}(0) = 1/2$, and $C_{001}(0) = \sqrt{3}/2$; i.e., the two modes are initially excited with different amplitudes whereas an atom is in the ground state. Other parameters are $ \Delta \omega _{a} = \Delta \omega _{b} = \Omega$, $m = 1$, and $\Omega _{Ra} = \Omega _{Rb} =\Omega _{R}$. }
\label{fig5}
\end{figure}

An atom, originally in its ground state, will get excited through resonant coupling to the EM field, as is obvious from Eq.~(\ref{sv-excited-modes}). The resulting dynamics of the averaged normalized energy of an atom $\left\langle \Psi \right\vert \hat{H}_{a}\left\vert \Psi \right\rangle /W$ and the numbers of quanta in mode $a$, $|C_{100}(t)|^2$ and mode $b$, $|C_{100}(t)|^2$  is shown in Fig.~\ref{fig5} for one generic set of initial conditions. Due to the presence of three coupled degrees of freedom, the evolution is more complicated than single-sinusoidal Rabi oscillations. Moreover, there is one particular choice of initial conditions,   $C_{010}(0) = - C_{100}(0) \frac{R_{a0}}{R_{bm}}$, which corresponds to $B = C = 0$ and $A = C_{100}(0)$, where the normalization condition gives $|C_{100}(0)|^2 = \left( 1 + \frac{|R_{a0}|^2}{|R_{bm}|^2} \right)^{-1}$. This gives the following state vector,
\begin{equation}
\left(
\begin{array}{c}
C_{001} \\
C_{100} \\
C_{010}%
\end{array}%
\right) = C_{100}(0) \left(
\begin{array}{c}
0 \\
e^{-i\int_{0}^{t}\omega _{100}\left( \tau \right) d\tau } \\
-\frac{R_{a0}}{R_{bm}} 
e^{-i\int_{0}^{t}\omega _{010}\left( \tau \right) d\tau }%
\end{array}%
\right)   .  \label{coe for one excited cav mode}
\end{equation}%

It corresponds to
the solution in which an atom stays in the ground state and \textit{is not excited by the electromagnetic
field despite being in resonance}. It happens because of destructive
interference between two frequency-modulated electromagnetic modes. In this case the three quantities shown in Fig.~\ref{fig5} become constant in time, with the average atomic energy being zero at all times. This effect is discussed in more detail in Sec.~V where the dissipation is taken into account.

%%%%%%%%%%%%%%%%%%%%%%%%%%%%%%%%%%%%%%%

\section{Dynamics of two cavity modes coupled to a time-variable atom}

Consider now the situation in which the cavity is not changing with time
whereas the transition energy of an atom depends on the parameter $p$ which
is adiabatically modulated. For example, it could be an optical transition
in a semiconductor nanostructure under an applied time-variable bias. The
Hamiltonian of such an atom can be written as $\hat{H}=W\left( p\right) \hat{%
\sigma}^{\dagger }\hat{\sigma}$. The dynamics of an isolated atom conserves
the adiabatic invariant $\frac{\left\langle \Psi \right\vert \hat{H}%
\left\vert \Psi \right\rangle }{\omega \left( t\right) }$ , where $\omega
\left( t\right) =\frac{W\left( p\left( t\right) \right) }{\hbar }$.

The dipole moment of the transition is also modulated, $\left\langle
1\right\vert \mathbf{\hat{d}}\left\vert 0\right\rangle =\mathbf{d}\left(
p\left( t\right) \right) $, because atom wave functions in the coordinate
representation depend on the parameter $p$. We again consider small enough amplitude of modulation of the transition energy. In this case, using the arguments
similar to those in Sec.~II B, we can show that the dependence $\mathbf{d}(t) $ can be neglected; it is the dependence $W(t) $
which is important for the evolution of a coupled atom-field system. The
RWA Hamiltonian which describes such a system is
\begin{equation}
\hat{H}=\hbar \omega _{a}\left( \hat{a}^{\dagger }\hat{a}+\frac{1}{2}%
\right) +\hbar \omega _{b}\left( \hat{b}^{\dagger }\hat{b}+\frac{1}{2%
}\right) +W\left( t\right) \hat{\sigma}^{\dagger }\hat{\sigma}-\left[ \hat{%
\sigma}^{\dagger }\left( \chi _{a}\hat{a}+\chi _{b}\hat{b}\right) +\hat{%
\sigma}\left( \chi _{a}^{\ast }\hat{a}^{\dagger }+\chi _{b}^{\ast }\hat{b}%
^{\dagger }\right) \right] .  \label{RWA H}
\end{equation}
Consider again a sinusoidal modulation of the transition energy,

\begin{equation}
W\left( t\right) =\bar{W}-\hbar \Delta \omega \sin \left( \Omega t\right) ,
\label{Wt}
\end{equation}%
where $\frac{\bar{W}}{\hbar }\gg \Delta \omega $.

The Schr\"{o}dinger equation with this Hamiltonian allows analytic
solutions. For simplicity, we again consider the basis states with lowest
energies: $\left\vert 0_{a}\right\rangle \left\vert 0_{b}\right\rangle
\left\vert 0\right\rangle$, $\left\vert 0_{a}\right\rangle \left\vert
0_{b}\right\rangle \left\vert 1\right\rangle$, $\left\vert 1_{a}\right\rangle
\left\vert 0_{b}\right\rangle \left\vert 0\right\rangle$, and $\left\vert
0_{a}\right\rangle \left\vert 1_{b}\right\rangle \left\vert 0\right\rangle$. 
The corresponding wave function is
\begin{equation}
\Psi =C_{000}\left\vert 0_{a}\right\rangle \left\vert 0_{b}\right\rangle
\left\vert 0\right\rangle +C_{001}\left\vert 0_{a}\right\rangle \left\vert
0_{b}\right\rangle \left\vert 1\right\rangle +C_{100}\left\vert
1_{a}\right\rangle \left\vert 0_{b}\right\rangle \left\vert 0\right\rangle
+C_{010}\left\vert 0_{a}\right\rangle \left\vert 1_{b}\right\rangle
\left\vert 0\right\rangle,
   \label{WF}
\end{equation}%
where the coefficients obey the equations%
\begin{equation}
\dot{C}_{000}+i\frac{\omega _{a}+\omega _{b}}{2}C_{000}=0;  \label{C000 dot}
\end{equation}%
\begin{equation}
\dot{C}_{001}+i\left( \frac{1}{2}\omega _{a}+\frac{1}{2}\omega _{b}+\frac{%
W\left( t\right) }{\hbar }\right) C_{001}-i\Omega _{Ra}C_{100}-i\Omega
_{Rb}C_{010}=0,  \label{C001 dot}
\end{equation}%
\begin{equation}
\dot{C}_{100}+i\left( \frac{3}{2}\omega _{a}+\frac{1}{2}\omega _{b}\right)
C_{100}-i\Omega _{Ra}^{\ast }C_{001}=0,  \label{C100 dot}
\end{equation}%
\begin{equation}
\dot{C}_{010}+i\left( \frac{1}{2}\omega _{a}+\frac{3}{2}\omega _{b}\right)
C_{010}-i\Omega _{Rb}^{\ast }C_{001}=0.
  \label{C010 dot}
\end{equation}%
After the substitution%
\begin{equation}
\left(
\begin{array}{c}
C_{001} \\
C_{100} \\
C_{010}%
\end{array}%
\right) =\left(
\begin{array}{c}
G_{0}\exp \left[ -i\left( \frac{1}{2}\omega _{a}+\frac{1}{2}\omega
_{b}\right) t+\int_{0}^{t}\frac{W\left( \tau \right) }{\hbar }d\tau \right]
\\
G_{a}\exp \left[ -i\left( \frac{3}{2}\omega _{a}+\frac{1}{2}\omega
_{b}\right) t\right]  \\
G_{b}\exp \left[ -i\left( \frac{1}{2}\omega _{a}+\frac{3}{2}\omega
_{b}\right) t\right]
\end{array}%
\right) ,  \label{Substitution}
\end{equation}%
we obtain%
\begin{equation}
\dot{G}_{0}-i\Omega _{Ra}G_{a}\exp \left[ -i\left( \omega _{a}t-\int_{0}^{t}%
\frac{W\left( \tau \right) }{\hbar }d\tau \right) \right] -i\Omega
_{Rb}G_{b}\exp \left[ -i\left( \omega _{b}t-\int_{0}^{t}\frac{W\left( \tau
\right) }{\hbar }d\tau \right) \right] =0,  \label{G0 dot}
\end{equation}%
\begin{equation}
\dot{G}_{a}-i\Omega _{Ra}^{\ast }G_{0}\exp \left[ -i\left( \omega
_{a}t-\int_{0}^{t}\frac{W\left( \tau \right) }{\hbar }d\tau \right) \right]
=0,  \label{Ga dot}
\end{equation}%
\begin{equation}
\dot{G}_{b}-i\Omega _{Rb}^{\ast }G_{0}\exp \left[ -i\left( \omega
_{b}t-\int_{0}^{t}\frac{W\left( \tau \right) }{\hbar }d\tau \right) \right]
=0,  \label{Gb dot}
\end{equation}%
Similarly to the previous section, we expand the exponents in Eqs.~(\ref{G0
dot})-(\ref{Gb dot}) over the harmonics of the modulation frequency $\Omega
$ using Eq.~(\ref{coe}) and keep only the resonant terms, assuming for
definiteness that $\omega _{a}=\frac{\bar{W}}{\hbar }$ and $\omega
_{b}+m\Omega =\frac{\bar{W}}{\hbar }$. We again obtain Eq.~(\ref{simp eq}), where $R_{a0} = \Omega_{Ra} J_0\left(\frac{\Delta \omega}{\Omega}\right)$, $R_{bm} = (-i)^{|m|}  \Omega _{Rb} J_{|m|}\left(\frac{\Delta \omega}{\Omega}\right)$. 
Therefore, the
modulation of the atomic transition and the cavity parameters leads to a
similar dynamics.

%%%%%%%%%%%%%%%%%%%%%%%%%%%%

\section{Dynamics of open time-dependent cavity QED systems}

\subsection{The stochastic evolution of the state vector}

Consider again the dynamics of two adiabatically varying cavity modes
coupled to an atom, but this time we include the processes of relaxation and decoherence in an open
system, which is (weakly) coupled to a dissipative reservoir. We will use the approach based on the 
stochastic evolution of the state vector; see Appendix B and
\cite{tokman2020}. 
This is basically the Schr\"{o}dinger equation modified by
adding a linear relaxation operator and the noise source term with
appropriate correlation properties. The latter are related to the parameters
of the relaxation operator, which is a manifestation of the
fluctuation-dissipation theorem \cite{Landau1965}. In Appendix B we outlined the main properties of the stochastic equation of evolution 
and showed how physically reasonable constraints on the observables
determine the properties of the noise sources. We also demonstrated the
relationship between our approach and the Lindblad method of solving the
master equation.

Within our approach the system is described by a state vector which has a
fluctuating component: $\left\vert \Psi \right\rangle =\overline{\left\vert
\Psi \right\rangle }+\widetilde{\left\vert \Psi \right\rangle }$, where the
straight bar means averaging over the statistics of noise and the wavy bar
denotes the fluctuating component. This state vector is of course very
different from the state vector obtained by solving a standard Schr\"{o}%
dinger equation for a closed system. In fact, coupling to a dissipative
reservoir leads to the formation of a mixed state, which can be described by
a density matrix $\hat{\rho}=\overline{\left\vert \Psi \right\rangle }\cdot
\overline{\left\langle \Psi \right\vert }+\overline{\widetilde{\left\vert
\Psi \right\rangle }\widetilde{\left\langle \Psi \right\vert }}$. However, the density matrix equations are more cumbersome for the analytic solution as compared to the formalism used in this paper. 

One can view the stochastic equation approach as a convenient formalism for
calculating physical observables which allows one to obtain analytic solutions for the evolution of a coupled system in the presence of dissipation and decoherence. When the Markov approximation is applied, the results are equivalent to those obtained within the Lindblad master equation formalism. Within the Markov approximation, the relaxation operator in the stochastic equation for the state vector is obtained simply by summing up partial Lindbladians for all subsystems, whatever they are (in our case these are a fermion emitter and two EM cavity modes). Then the noise source term is determined unambiguously by conservation of the norm of the state vector and the requirement that the system should approach thermal equilibrium when the external perturbation is turned off. This immediately gives Eqs.~(\ref{modified c000 dot})-(\ref{modified c010 dot}) below.

Following the derivation in Appendix B, equations~(\ref{c000 dot})-(\ref{c010 dot}) are modified due
to the terms with relaxation constants $\gamma _{000}$,$\gamma _{001}$,$\gamma _{010}$, and $\gamma _{100}$ which are originated from the Lindladians,  and
the noise sources,
\begin{equation}
\left( \frac{\partial }{\partial t}+\gamma _{000}\right) C_{000}+i\frac{%
\omega _{a}\left( t\right) +\omega _{b}\left( t\right) }{2}C_{000}=-\frac{i}{%
\hbar }\mathfrak{R}_{000};  \label{modified c000 dot}
\end{equation}%
\begin{equation}
\left( \frac{\partial }{\partial t}+\gamma _{001}\right) C_{001}+i\left(
\frac{1}{2}\omega _{a}\left( t\right) +\frac{1}{2}\omega _{b}\left( t\right)
+\frac{W}{\hbar }\right) C_{001}-i\Omega _{Ra}C_{100}-i\Omega _{Rb}C_{010}=-%
\frac{i}{\hbar }\mathfrak{R}_{001},  \label{modified c001 dot}
\end{equation}%
\begin{equation}
\left( \frac{\partial }{\partial t}+\gamma _{100}\right) C_{100}+i\left(
\frac{3}{2}\omega _{a}\left( t\right) +\frac{1}{2}\omega _{b}\left( t\right)
\right) C_{100}-i\Omega _{Ra}^{\ast }C_{001}=-\frac{i}{\hbar }\mathfrak{R}_{100},
\label{modified c100 dot}
\end{equation}%
\begin{equation}
\left( \frac{\partial }{\partial t}+\gamma _{010}\right) C_{010}+i\left(
\frac{1}{2}\omega _{a}\left( t\right) +\frac{3}{2}\omega _{b}\left( t\right)
\right) C_{010}-i\Omega _{Rb}^{\ast }C_{001}=-\frac{i}{\hbar }\mathfrak{R}_{010}.
\label{modified c010 dot}
\end{equation}%

We assume that  noise terms in Eq.~(\ref{modified c000 dot})-(\ref{modified c010 dot}) become equal to zero after averaging over the noise statistics. The averages of the
quadratic combinations of noise source terms are nonzero and we assume here that they are delta-correlated in time (the Markov approximation),  
\begin{equation}
\overline{\mathfrak{R}_{\beta }^{\ast }\left( t+\xi \right) \mathfrak{R}_{\alpha }\left( t\right) }%
=\overline{\mathfrak{R}_{\beta }^{\ast }\left( t\right) \mathfrak{R}_{\alpha }\left( t+\xi \right)
}=\hbar ^{2}\delta \left( \xi \right) D_{\alpha \beta }.  \label{correlator}
\end{equation}%
Here the indices $\alpha$ and $\beta$ span a set of the lowest-energy states $\left\vert 0_{a}\right\rangle \left\vert 0_{b}\right\rangle
\left\vert 0\right\rangle$, $\left\vert 0_{a}\right\rangle \left\vert
0_{b}\right\rangle \left\vert 1\right\rangle$, $\left\vert 1_{a}\right\rangle
\left\vert 0_{b}\right\rangle \left\vert 0\right\rangle$, and $\left\vert
0_{a}\right\rangle \left\vert 1_{b}\right\rangle \left\vert 0\right\rangle$.  Including the
noise sources is crucial for consistency of the formalism: it ensures the
conservation of the norm of the state vector and leads to a physically
meaningful equilibrium state. 

Consider the case of zero temperatures for all reservoirs, which means in
practice that these temperatures in energy units are much lower than the
atomic transition energy and the cavity mode frequencies. In this case the relaxation constants are greatly simplified as compared to the general expressions given in Appendix B, 
\begin{equation}
\gamma_{000} = 0, \, \gamma_{001} = \frac{\gamma }{2}, \, 
\gamma _{100}=\frac{\mu _{a}}{2}, \,  
\gamma _{010}= \frac{\mu
_{b}}{2},  \label{gamma at 0 temp}
\end{equation}
where $\gamma $ is the inelastic relaxation rate for an isolated atom, $\mu _{a,b}$
are relaxation rates of the EM modes determined by the cavity Q-factor; these ``partial'' relaxation constants are determined by couplings to their respective dissipative reservoirs. Appendix B outlines how to include elastic decoherence processes.

In this limit we can drop the noise terms in the right-hand side of all
equations for the components of the state vector, except the term $\mathfrak{R}_{000}$
in the equation for $C_{000}$; see Appendix B. This noise term ensures
conservation of the norm,
\begin{equation*}
 \overline{
\left\vert C_{000}\right\vert ^{2}}+\overline{\left\vert
C_{001}\right\vert ^{2}}+ \overline{\left\vert
C_{010}\right\vert ^{2}}+ \overline{\left\vert
C_{100}\right\vert ^{2}} =0,
\end{equation*}%
if its correlator is given by 
\begin{equation*}
\overline{\mathfrak{R}_{000}\left( t+\xi \right) \mathfrak{R}_{000}^{\ast }\left( t\right) }%
= 2\hbar ^{2}\delta \left( \xi \right) \left( \gamma _{100}\overline{\left\vert C_{100}\right\vert
^{2}} +\gamma _{001}\overline{
\left\vert C_{001}\right\vert ^{2}} +\gamma _{010}\overline{
\left\vert C_{010}\right\vert ^{2}} \right).
\end{equation*}

As an example, consider a high-quality cavity and neglect the cavity losses as compared to the atomic decay.
In this case, and for a low temperature of an atomic reservoir, Eqs.~(\ref%
{modified c001 dot})-(\ref{modified c010 dot}) take the form%
\begin{equation}
\left( \frac{\partial }{\partial t}+\frac{\gamma }{2}\right) C_{001}+i\left(
\frac{1}{2}\omega _{a}\left( t\right) +\frac{1}{2}\omega _{b}\left( t\right)
+\frac{W}{\hbar }\right) C_{001}-i\Omega _{Ra}C_{100}-i\Omega _{Rb}C_{010}=0,
\label{modified c001 dot at low temp}
\end{equation}%
\begin{equation}
\frac{\partial }{\partial t}C_{100}+i\left( \frac{3}{2}\omega _{a}\left(
t\right) +\frac{1}{2}\omega _{b}\left( t\right) \right) C_{100}-i\Omega
_{Ra}^{\ast }C_{001}=0,  \label{modified c100 dot at low temp}
\end{equation}%
\begin{equation}
\frac{\partial }{\partial t}C_{010}+i\left( \frac{1}{2}\omega _{a}\left(
t\right) +\frac{3}{2}\omega _{b}\left( t\right) \right) C_{010}-i\Omega
_{Rb}^{\ast }C_{001}=0.  \label{modified c010 dot at low temp}
\end{equation}%
Using the substitution of variables in Eq.~(\ref{substitution}) and
repeating the same derivation as in Sec.~III, we arrive at 
\begin{equation}
\frac{d}{dt}\left(
\begin{array}{c}
G_{0} \\
G_{a} \\
G_{b}%
\end{array}%
\right) +\left(
\begin{array}{ccc}
\frac{\gamma}{2} & -i R_{a0} &
-i R_{bm} \\
-iR_{a0}^{\ast } & 0 & 0 \\
-i R_{bm}^{\ast} & 0 & 0
\end{array}%
\right) \left(
\begin{array}{c}
G_{0} \\
G_{a} \\
G_{b}%
\end{array}%
\right) =0.
\label{modified simp eq}
\end{equation}%
Its solution is determined by the eigenvalues and eigenvectors of the matrix
in Eq.~(\ref{modified simp eq}). The eigenvalues are given by%
\begin{equation*}
\Gamma \left[ \left( \Gamma -\frac{\gamma }{2}\right) \Gamma +\Omega
_{R\Sigma }^{2}\right] =0,
\end{equation*}%
which yields%
\begin{equation}
\Gamma _{0}=0,\ \Gamma _{1,2}=\frac{\gamma }{4}\pm i\sqrt{\Omega _{R\Sigma
}^{2}-\frac{\gamma ^{2}}{16}}.  \label{modified eigenvalues}
\end{equation}%
The eigenvector corresponding to the eigenvalue $\Gamma _{0}=0$ is the same
as in the absence of dissipation (see Sec.~III B), whereas the expressions for
the eigenvectors corresponding to eigenvalues $\Gamma _{1,2}$ can be
obtained from \textquotedblleft dissipationless\textquotedblright\
expressions by replacing $\pm $ $\Omega _{R\Sigma }\longrightarrow \pm \sqrt{%
\Omega _{R\Sigma }^{2}-\frac{\gamma ^{2}}{16}}-i\frac{\gamma }{4}$. As a
result, we obtain the following expression for the state vector,%
\begin{eqnarray}
\left(
\begin{array}{c}
C_{001} \\
C_{100} \\
C_{010}%
\end{array}%
\right) &=&A\left(
\begin{array}{c}
0 \\
e^{-i\int_{0}^{t}\omega _{100}\left( \tau \right) d\tau } \\
- \frac{R_{a0}}{R_{bm}} 
e^{-i\int_{0}^{t}\omega _{010}\left( \tau \right) d\tau }%
\end{array}%
\right)  \notag \\
+ &&B e^{\left( -i\sqrt{\Omega _{R\Sigma }^{2}-\frac{\gamma ^{2}}{16}}-\frac{%
\gamma }{4}\right) t}\left(
\begin{array}{c}
\frac{\sqrt{\Omega _{R\Sigma }^{2}-\frac{\gamma ^{2}}{16}}-i\frac{\gamma }{4}%
}{R_{a0}^{\ast} }%
e^{-i\int_{0}^{t}\omega _{001}\left( \tau \right) d\tau } \\
e^{-i\int_{0}^{t}\omega _{100}\left( \tau \right) d\tau } \\
 \frac{R_{bm}^{\ast}}{R_{a0}^{\ast}}  e^{-i\int_{0}^{t}\omega _{010}\left( \tau \right) d\tau }%
\end{array}%
\right)  \notag \\
&& + C e^{\left( i\sqrt{\Omega _{R\Sigma }^{2}-\frac{\gamma ^{2}}{16}}-\frac{%
\gamma }{4}\right) t}\left(
\begin{array}{c}
\frac{-\sqrt{\Omega _{R\Sigma }^{2}-\frac{\gamma ^{2}}{16}}-i\frac{\gamma }{4%
}}{R_{a0}^{\ast}}  e^{-i\int_{0}^{t}\omega _{001}\left( \tau \right) d\tau } \\
e^{-i\int_{0}^{t}\omega _{100}\left( \tau \right) d\tau } \\
 \frac{R_{bm}^{\ast}}{R_{a0}^{\ast}}  e^{-i\int_{0}^{t}\omega _{010}\left( \tau \right) d\tau }%
\end{array}%
\right) .  \label{sol for modified simp eq}
\end{eqnarray}%
Where the constants $A$, $B$ and $C$ are given by initial conditions. In the limit $\Omega_{R\Sigma} \gg \gamma$ their dependence on the initial values $C_{100}(0)$, $C_{010}(0)$, and $C_{001}(0)$ is given by Eqs.~(\ref{abc}) from the previous section,  whereas their dependence on $C_{000}(0)$ is determined by the normalization condition. 

%%%%%%%%%%%%%%%%%%%%%%%%%%%%%%%%%%%%%%%%%%%%%

\subsection{Modulation-induced transparency}

Note
again the existence of the solution with $B = C = 0$ in which an atom initially in the ground state \textit{is decoupled
from the electromagnetic field} and stays in the ground state because of destructive interference between the EM modes. 
There is however an interesting difference as compared to the dissipationless case discussed in Sec.~IIE. For arbitrary initial conditions, when $A,B,C$ are not equal to zero, part of the field energy will be resonantly transferred to the atom and dissipate through the atomic decay. However, the terms with $B$ and $C$ factors in Eq.~(\ref{sol for modified simp eq}) decay exponentially with time, and  the solution to Eq.~(\ref{sol for modified simp eq}) at $ t \gg 1/\gamma$ will acquire the same form as  in the case of $B = C = 0$:
\begin{align}
\Psi & =A \left( e^{-i\int_{0}^{t}\omega _{100}\left( \tau \right)
d\tau }\left\vert 1_{a}\right\rangle \left\vert 0_{b}\right\rangle
\left\vert 0\right\rangle  - \frac{R_{a0}}{R_{bm}}   e^{-i\int_{0}^{t}\omega _{010%
}\left( \tau \right) d\tau }\left\vert 0_{a}\right\rangle \left\vert
1_{b}\right\rangle \left\vert 0\right\rangle \right)  +C_{000}\left\vert
0_{a}\right\rangle \left\vert 0_{b}\right\rangle \left\vert 0\right\rangle
\label{modified wf}. 
\end{align}

The value of $C_{000}$ at $ t \gg 1/\gamma$ is determined by the noise term $-\frac{i}{\hbar} \mathfrak{R}_{000}$ in the right-hand side of Eq.~(\ref{modified c000 dot}) and satisfies $\overline{C_{000}} = 0$, $\overline{|C_{000}|^2} = 1- |A|^2 \left[ 1 + \frac{|R_{a0}|^2}{|R_{bm}|^2} \right]$ (see Appendix B). 

The value of $|A|^2$ is given by 
\begin{equation}
    |A|^2 = \displaystyle \frac{1-|C_{000}(0)|^2 - |C_{001}(0)|^2}{1+|Z|^2} \left[ \frac{\left| \frac{|R_{bm}|^2}{|R_{a0}|^2} - Z \frac{R_{bm}}{R_{a0}} \right|^2 }{\left( 1+ \frac{|R_{bm}|^2}{|R_{a0}|^2} \right)^2 } \right],
\end{equation}
where $Z = \displaystyle \frac{C_{010}(0)}{C_{100}(0)}$. 
The value of $|A|^2$ reaches a maximum when Arg$[Z] = \pi - {\rm Arg}\left[\frac{R_{bm}}{R_{a0}} \right]$ and $|Z| = \left| \frac{R_{a0}}{R_{bm}} \right|$, which corresponds to $C_{010}(0) = - C_{100}(0) \frac{R_{a0}}{R_{bm}}$ and 
\begin{equation}
    \label{a2}
|A|^2 = \displaystyle \frac{1-|C_{000}(0)|^2 - |C_{001}(0)|^2}{1+ \frac{|R_{a0}|^2}{|R_{bm}|^2}}. 
\end{equation}
This equation has a simple interpretation. According to Eq.~(\ref{modified wf}), the average steady-state number of quanta in both modes is 
\begin{equation}
|C_{100}|^2 + |C_{010}|^2 = \displaystyle  |A|^2 \left(1+ \frac{|R_{a0}|^2}{|R_{bm}|^2} \right).
\label{avnumber}
\end{equation}
Comparing Eq.~(\ref{avnumber}) and Eq.~(\ref{a2}), one can see that despite the presence of dissipation, when the value of $|A|^2$ reaches a maximum given by Eq.~(\ref{a2}) the average steady-state number of field quanta given by Eq.~(\ref{avnumber}) is equal to its initial value:  $ |C_{100}(0)|^2 + |C_{010}(0)|^2 = 1-|C_{000}(0)|^2 - |C_{001}(0)|^2 $.

\begin{figure}[htb]
	\includegraphics[width=0.5\textwidth]{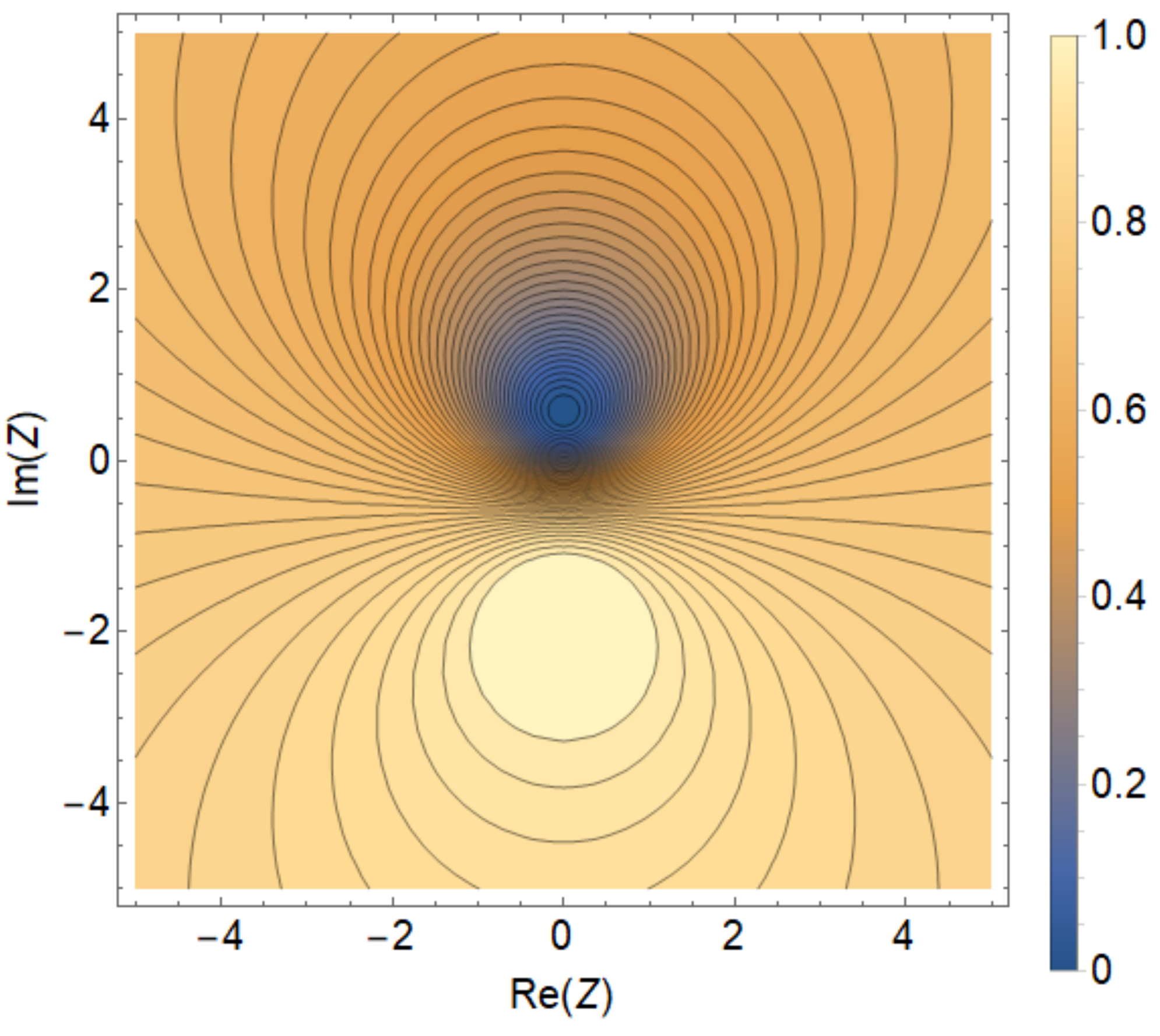}
\caption { The contour plot of the normalized average number of quanta $\overline{N_q}$ on the complex $Z$ plane for $m = 1$, $\Omega_{Ra} = \Omega_{Rb}$ and $\Delta \omega_a = \Delta \omega_b = \Omega$.}
\label{asquared}
\end{figure}

The contour plot of the  average steady-state number of quanta normalized by its initial value, 
$$ \overline{N_q} = \displaystyle  \frac{|A|^2 \left(1+ \frac{|R_{a0}|^2}{|R_{bm}|^2} \right)}{ |C_{100}(0)|^2 + |C_{010}(0)|^2  }$$ 
on the complex $Z$ plane is shown in Fig.~\ref{asquared} for $m = 1$, $\Omega_{Ra} = \Omega_{Rb}$ and $\Delta \omega_a = \Delta \omega_b = \Omega$. For this particular choice of parameters, the maximum of  the number of quanta is reached at Arg$[Z] = -\pi/2$, i.e., it is located on the imaginary axis as shown in the figure. For $m = 2$ the maximum will be on the real axis. At its maximum, the average number of quanta is equal to its initial value, i.e. it remains constant.

For any initial conditions other than those corresponding to the maximum of $\overline{N_q}$, a part of the EM field energy will dissipate through interaction with an atom, and eventually only the part which corresponds to the combination of modes completely decoupled from an atom due to destructive interference survives. This will result in smaller values of $\overline{N_q}$.  Of course, eventually the finite cavity losses will kick in and the
field will dissipate to the level of quantum and thermal fluctuations.

Finally, for the initial state $\Psi
\left( 0\right) =\left\vert 0_{a}\right\rangle \left\vert 0_{b}\right\rangle
\left\vert 1\right\rangle $ (only the atom is excited) we have $A=0$, i.e.
the system goes into the ground state as expected. 

To summarize, the modulated system of an atom resonantly coupled to two EM
cavity modes demonstrates an interesting effect of modulation-induced
transparency. In the absence of modulation, the presence of an atom
experiencing an incoherent decay leads to the dissipation of the EM field
 even if the empty cavity is ideal,
i.e. has zero losses. However, low-frequency modulation of the cavity or of
the transition frequency of an atom creates the EM field distribution which
is completely decoupled from an atom due to destructive interference between
the cavity modes, even at resonance between the atomic transition and the
cavity mode frequencies. Therefore, the atom will remain in the ground state and the field will experience no
dissipation in the absence of cavity losses. For a classical field, such a destructive interference effect which switches off the
field dissipation in resonant medium by introducing low-frequency modulation
was considered, in particular, in Ref.~\cite{rad2006} for acoustically modulated
two-level atoms. Similar effects in the interaction of classical fields with atoms are discussed in the introduction of Ref.~\cite{rad2020}. 

\subsection{Prospects for strong coupling and quantum entanglement in various nanophotonic systems}

Expressions in this section and more general expressions for the relaxation rates in Appendix B (see, e.g., Eqs.~(\ref{gamma0}),(\ref{gamma1})) allow one to calculate the effective decoherence rates from the known  ``partial'' relaxation rates
for individual subsystems: EM cavity modes and any kind of a fermionic qubit. One can compare the decoherence rates with characteristic Rabi frequencies which enter the solution for the evolution equations such as Eqs.~(\ref{modified c001 dot at low temp})-(\ref{modified c010 dot at low temp}) in order to determine if the strong coupling regime and quantum entanglement in the electron-photon system can be achieved. For any specific application, one should also compare the effective relaxation times with relevant operation times (gate transition time, read/write time etc.) In the discussion below, we rely on the parameters obtained from Refs.~\cite{haroche}-\cite{aspelmeyer2014} which we already cited in the Introduction. Many of them are recent reviews and one can find further references there. We don't attempt to overview here the vast and rapidly growing amount of literature on the subject.

In electron-based quantum emitters the largest oscillator strengths in the visible/near-infrared range have been observed for excitons in organic molecules, followed by perovskites and more conventional inorganic semiconductor quantum dots. The typical variation of the dipole matrix element of the optical transition which enters the Rabi frequency is from tens of nm to a few Angstrom. The dipole moment grows with increasing wavelength. The relaxation times are strongly temperature and material quality-dependent, varying from tens or hundreds of ps for single quantum dots at 4 K to the $\mu$s range for defects in semiconductors and diamond at mK temperatures. At room temperature the typical decoherence rates for the optical transition are in the $\sim 10$ meV range. 

The photon decay times are longest for dielectric micro- and nano-cavities: photonic crystal cavities, nanopillars, distributed Bragg reflector mirrors, microdisk whispering gallery mode cavities, etc. Their quality factors are typically between $10^3 - 10^7$, corresponding to photon lifetimes from sub-ns to $\mu$s range. However, the field localization in the dielectric cavities is diffraction-limited, which limits the attainable Rabi frequency values to hundreds of $\mu$eV. The effective decoherence rate in dielectric cavity QED systems is typically limited by the relaxation in the fermion quantum emitter subsystem,  

In plasmonic cavities, field localization on a nm and even sub-nm scale has been achieved, but the photon losses are in the ps or even fs range and therefore, they dominate the overall decoherence rate. Still, when it comes to strong coupling at room temperature to a single quantum emitter such as a single molecule or a quantum dot, the approach utilizing plasmonic cavities has seen more success so far. In these systems the Rabi splitting of the order of 100-200 meV has been observed. In plasmonic systems it may be beneficial to consider longer-wavelength emitters with the optical transition at the mid-infrared and even terahertz wavelengths. Indeed, with increasing wavelength the plasmon losses go down, the matrix element of a dipole-allowed transition increases, whereas the plasmon localization stays largely the same. 

Another factor that has to be taken into account when choosing a nanophotonic system for a specific application is the rate with which the modulation of the cavity or emitter parameters has to be performed. For example, if the modulation at the rate comparable to the Rabi frequency or operation with $\pi$- or $\pi/2$ pulses is required, the plasmonic-based systems run into a problem: they would require $\sim 10-100$ fs pulses for modulation, which obviously can be achieved only with fs lasers. All electronic operations typically have a cutoff at tens of GHz. Applications of nanophotonics to quantum computing are especially challenging, because  computations require at least 99.99\% fidelity, i.e. at least $10^4$ ``flops'' before decoherence kicks in.

%%%%%%%%%%%%%%%%%%%%%%%%%%%%

\section{Conclusions} 

In conclusion, we developed the analytic theory describing the dynamics and control  of strongly coupled nanophotonic systems with time-variable parameters. The coupling of the fermion and photon subsystems to their dissipative reservoirs are  included within the stochastic equation of evolution approach, which is equivalent to the Lindblad approximation in the master equation formalism. Our analytic solution is valid in the approximation that the rate of parameter modulation and the amplitude of the frequency modulation are much smaller than the optical transition frequencies. At the same time, they can be arbitrary with respect to the generalized Rabi oscillations frequency which determines the coherent dynamics. Therefore, we can describe an arbitrary modulation of the parameters, both slower and faster than the Rabi frequency, for complete control of the quantum state. For example, one can turn on and off the entanglement between the fermionic and photonic degrees of freedom, swap between the quantum states, or decouple  the fermionic qubit from the cavity field via modulation-induced transparency.

\begin{acknowledgments}
This work has been supported in part by the Air Force Office for Scientific Research 
Grant No.~FA9550-17-1-0341, National Science Foundation Award No.~1936276, and Texas A\&M University through STRP, X-grant and T3-grant programs.
 M.E.~and M.T.~acknowledge the support from RFBR Grant No. 20-02-00100.

\end{acknowledgments}

%%%%%%%%%%%%%%%%%%%%%%%%%%%%%
\appendix
%\renewcommand{\appendixname}{Appendix~\Alph{section}}

%%%%%%%%%%%%%%%%%%%%%%%%%%%%%%%%%

\section{Quantization of a cavity surface plasmon field}

Consider a planar cavity oriented parallel to $(x,y)$ plane and sandwiched
between two layers of material with isotropic dielectric constant $%
\varepsilon \left( \omega \right) $ which could be dielectric or metal. The
transverse size of a cavity along $z$ is from $z=-d$ to $z=+d$. The
dielectric constant inside the cavity is $\varepsilon _{g}\left( \omega
\right) $, also assumed isotropic.

\subsection{Spatial structure of the field and frequency dispersion}

Whether or not the field is quantized, its distribution in space and
frequencies of modes are determined from solving the boundary value problem
of classical electrodynamics. Here we consider the field localized to a
subwavelength region, to scales $l_{SP}\ll \frac{c}{\varepsilon _{g}\left(
\omega \right) \omega },\frac{c}{\left\vert \varepsilon \left( \omega
\right) \right\vert \omega }$ , which allows us to use electrostatic
approximation. We seek the solution for the electric potential as $\varphi
=\Phi \left( z\right) e^{i\mathbf{k}\cdot \mathbf{r-}i\omega t}$ , where the
2D vectors $\mathbf{r,k}$ are in the $x,y$ plane. The Poisson's equation for
the potential in every region has a form%
\begin{equation}
\frac{\partial ^{2}\Phi }{\partial z^{2}}=k^{2}\Phi  \label{Poisson equ}
\end{equation}

In the region $z<-d$ the solution is $\Phi =\Phi _{-}e^{kz}$ , whereas in $%
z>d$ the solution is $\Phi =\Phi _{+}e^{-kz}$.

Since the cavity is symmetric with respect to $z=0$, the spatial
distribution inside the cavity can be either symmetric,
$\Phi =\Phi _{s}\cosh \left( kz\right)  \label{sym}$, 
or antisymmetric, $ \Phi =\Phi _{as}\sinh \left( kz\right)  \label{asym}$.

The boundary conditions include the continuity of the potential and the $z$%
-component of the electric induction. 

\textbf{(i) Symmetric solution}: $\Phi _{-}=\Phi _{+}$. Substituting $z=-d$
the boundary conditions give%
\begin{equation}
\tanh \left( kd\right) =-\frac{\varepsilon }{\varepsilon _{g}}.  \label{bc}
\end{equation}%
i.e. we always need $\varepsilon \left( \omega \right) <0$ for positive $%
\varepsilon _{g}$. In the limit $kd\gg 1$ , Eq.~(\ref{bc}) corresponds to
the dispersion equation for a surface plasmon at the boundary between the
two infinite media
\begin{equation}
1=-\frac{\varepsilon }{\varepsilon _{g}},  \label{de}
\end{equation}%
whereas in the opposite limit $kd\longrightarrow 0$ and assuming that $%
\varepsilon _{g}$ is positive and not too small, we obtain a standard
dispersion equation for a plasmon in the bulk medium:\ $\varepsilon \left(
\omega \right) =0$.

Therefore, when $kd$ changes from $0$ to $\infty $ the symmetric surface
plasmon exists within a frequency bandwidth determined by the variation of $%
\frac{\varepsilon }{\varepsilon _{g}}$ from $-0$ to $-1$.

\textbf{\ (ii) Antisymmetric solution:}\ $\Phi _{-}=-\Phi _{+}$. The
boundary conditions give

\begin{equation}
\coth \left( kd\right) =-\frac{\varepsilon }{\varepsilon _{g}}.  \label{BC}
\end{equation}%
i.e. again $\varepsilon \left( \omega \right) <0$ for positive $\varepsilon
_{g}$ .

In the limit of a wide cavity, when $kd\gg 1$ the solution should
again corresponds to the surface plasmon at the boundary between the two
infinite media, i.e. we arrive at Eq.~(\ref{de}).

In the opposite limit $kd\longrightarrow 0$ and assuming that $%
\varepsilon _{g}$ is positive and not too small, we obtain that $\varepsilon
\left( \omega \right) \longrightarrow -\infty $. Therefore, when $kd$
changes from $0$ to $\infty $ the antisymmetric surface plasmon exists within a
frequency bandwidth determined by the variation of $\frac{\varepsilon \left(
\omega \right) }{\varepsilon _{g}\left( \omega \right) }$ from $-\infty $ to
$-1$.

 Note that in any case the electrostatic solution requires that $k\gg
\frac{\varepsilon _{g}\omega }{c},\frac{\left\vert \varepsilon \right\vert
\omega }{c}$.

\subsection{Field quantization}

Following \cite{Tokman2016}, we consider a cylinder with an axis of symmetry
along $z$ (i.e. orthogonal to the boundaries) and area $S$ in the $x,y$
plane. We assume that the field goes to $0$ when $z\longrightarrow \pm
\infty $ and satisfies periodic boundary conditions at the side surface of
the cylinder:

\begin{equation}
\mathbf{\hat{E}}=\sum_{\mathbf{k,}p}\mathbf{E}_{\mathbf{k,}p}\left( z\right)
e^{i\mathbf{k}\cdot \mathbf{r-}i\omega _{_{\mathbf{k,}p}}t}\hat{c}_{\mathbf{%
k,}p}+H_{.}c_{.},  \label{Ef}
\end{equation}%
where $p=s,as$ .

The spatial distribution of the field $\mathbf{E}_{\mathbf{k,}p}\left(
z\right) e^{i\mathbf{k}\cdot \mathbf{r}}$ and its frequency $\omega _{_{%
\mathbf{k,}p}}$ are given by the solution of the classical boundary value
problem in the previous section. The Hamiltonian $\hat{H}=\hbar \sum_{%
\mathbf{k,}p}\omega _{\mathbf{k,}p}\left( \hat{c}_{\mathbf{k,}p}^{\dagger }%
\hat{c}_{\mathbf{k,}p}+\frac{1}{2}\right) $ can be obtained from the
normalization condition \cite{Tokman2016}:

\begin{equation}
S\int_{-\infty }^{\infty }\left( \frac{\partial \left[ \omega \varepsilon
\left( \omega ,z\right) \right] }{\partial \omega }\mathbf{E}_{\mathbf{k,}%
p}^{\ast }\left( z\right) \mathbf{E}_{\mathbf{k,}p}\left( z\right) +\mathbf{B%
}_{\mathbf{k,}p}^{\ast }\left( z\right) \mathbf{B}_{\mathbf{k,}p}\left(
z\right) \right) dz=4\pi \hbar \omega _{_{\mathbf{k,}p}}  \label{NC-app}
\end{equation}%
where $S\int_{-\infty }^{\infty }\left( \cdot \cdot \cdot \right) dz=$ $%
\int_{V}\left( \cdot \cdot \cdot \right) dV$. For periodic or
\textquotedblleft cavity\textquotedblright\ boundary conditions we always
have \cite{Tokman2016}:%
\begin{equation}
\int_{V}\mathbf{B}_{\mathbf{k,}p}^{\ast }\mathbf{B}_{\mathbf{k,}%
p}dV=\int_{V}\varepsilon \mathbf{E}_{\mathbf{k,}p}^{\ast }\mathbf{E}_{%
\mathbf{k,}p}dV,  \label{pbc}
\end{equation}%
Which allows us to rewrite Eq.~(\ref{NC-app}) as
\begin{equation}
S\int_{-\infty }^{\infty }\frac{\partial \left[ \omega ^{2}\varepsilon
\left( \omega ,z\right) \right] }{\omega \partial \omega }\mathbf{E}_{%
\mathbf{k,}p}^{\ast }\left( z\right) \mathbf{E}_{\mathbf{k,}p}\left(
z\right) dz=4\pi \hbar \omega _{_{\mathbf{k,}p}}.  \label{norcon}
\end{equation}

For the fields $\mathbf{E}_{\mathbf{k,}p}\left( z\right) $ obtained in the
electrostatic approximation, we always obtain $\int_{V}\varepsilon \mathbf{E}%
_{\mathbf{k,}p}^{\ast }\mathbf{E}_{\mathbf{k,}p}dV=0$ , since in this
approximation $\mathbf{B}_{\mathbf{k,}p}=0$. In this case we can use the
normalization in the electrostatic limit:%
\begin{equation}
S\int_{-\infty }^{\infty }\frac{\partial \left[ \omega \varepsilon \left(
\omega ,z\right) \right] }{\partial \omega }\mathbf{E}_{\mathbf{k,}p}^{\ast
}\left( z\right) \mathbf{E}_{\mathbf{k,}p}\left( z\right) dz=4\pi \hbar
\omega _{_{\mathbf{k,}p}}.  \label{bcel}
\end{equation}

As a result, we obtain:

\textbf{(i) Symmetric mode} ($p=s$). The normalization condition:

\begin{equation}
S\left\vert \Phi _{s}\right\vert ^{2}k\left[ \frac{\partial \left( \omega
\varepsilon _{g}\right) }{\partial \omega }\sinh \left( 2kd\right) +2\cosh
^{2}\left( kd\right) \frac{\partial \left( \omega \varepsilon \right) }{%
\partial \omega }\right] =4\pi \hbar \omega _{_{\mathbf{k,}s}}  \label{symnc}
\end{equation}

\textbf{(ii)Antisymmetric mode }($p=as$). The normalization condition:

\begin{equation}
S\left\vert \Phi _{as}\right\vert ^{2}k\left[ \frac{\partial \left( \omega
\varepsilon _{g}\right) }{\partial \omega }\sinh \left( 2kd\right) +2\sinh
^{2}\left( kd\right) \frac{\partial \left( \omega \varepsilon \right) }{%
\partial \omega }\right] =4\pi \hbar \omega _{_{\mathbf{k,}as}}
\label{asymnc}
\end{equation}

Taking for simplicity $\varepsilon _{g}=1$ (air) and $\varepsilon \left(
\omega \right) =1-\frac{\omega _{pl}^{2}}{\omega ^{2}}$ (Drude dispersion)
gives

(\textbf{i) Symmetric mode} ($p=s$):

\begin{equation}
S\left\vert \Phi _{s}\right\vert ^{2}k\left[ 2\sinh \left( 2kd\right)
+4\cosh ^{2}\left( kd\right) \right] =4\pi \hbar \omega _{_{\mathbf{k,}s}}
\label{symNC}
\end{equation}

\textbf{(ii)Antisymmetric mode }($p=as$):%
\begin{equation}
S\left\vert \Phi _{as}\right\vert ^{2}k\left[ 2\sinh \left( 2kd\right)
+4\sinh ^{2}\left( kd\right) \right] =4\pi \hbar \omega _{_{\mathbf{k,}as}}
\label{asymNC}
\end{equation}

In order to calculate the coupling strength, it is important to know the
magnitude of the normalization field $\mathbf{E}_{\mathbf{k,}p}$ at the cavity boundary. Introducing the notation $\mathbf{E}_{\mathbf{k,}p}\left(
-d\right) =\mathbf{\tilde{E}}_{\mathbf{k,}p}$ and taking into account Eqs.~(%
\ref{sym}),(\ref{asym}),(\ref{symNC})~and~(\ref{asymNC}), we obtain%
\begin{equation}
\mathbf{\tilde{E}}_{\mathbf{k,}s}=\left[ \mathbf{z}_{0}k\sinh \left(
kd\right) -i\mathbf{k}\cosh \left( kd\right) \right] \sqrt{\frac{4\pi \hbar
\omega _{_{\mathbf{k,}s}}}{Sk\left[ 2\sinh \left( 2kd\right) +4\cosh
^{2}\left( kd\right) \right] }},  \label{norefs}
\end{equation}%
where%
\begin{equation}
\omega _{_{\mathbf{k,}s}}=\frac{\omega _{pl}}{\sqrt{1+\tanh \left( kd\right)
}};  \label{omegaks}
\end{equation}%
\begin{equation}
\mathbf{\tilde{E}}_{\mathbf{k,}as}=\left[ -\mathbf{z}_{0}k\cosh \left(
kd\right) -i\mathbf{k}\sinh \left( kd\right) \right] \sqrt{\frac{4\pi \hbar
\omega _{_{\mathbf{k,}as}}}{Sk\left[ 2\sinh \left( 2kd\right) +4\cosh
^{2}\left( kd\right) \right] }},  \label{norefas}
\end{equation}%
where
\begin{equation}
\omega _{_{\mathbf{k,}as}}=\frac{\omega _{pl}}{\sqrt{1+\coth \left(
kd\right) }}.  \label{omegakas}
\end{equation}

Figure 7 shows an example of normalized frequencies and field amplitudes of the symmetric and antisymmetric cavity modes given by Eqs.~(\ref{norefs})-(\ref{omegakas})  as a function of normalized time $\Omega t$ when the cavity height $d$ is modulated as $d(t) = d_0 (1 + 0.1 \sin(\Omega t))$. In this example $kd_0 = 1$. Even though the dependence of frequencies and field amplitudes on $kd$ is strongly nonlinear, their  modulation amplitudes remain small.  

%%%%%%%%%%%%%%%%%%%%%%%%%%%%%%%%%%%%

\begin{figure}[htb]
	\centering
	\begin{subfigure}{0.45\textwidth}
		\centering
		\includegraphics[width=\linewidth]{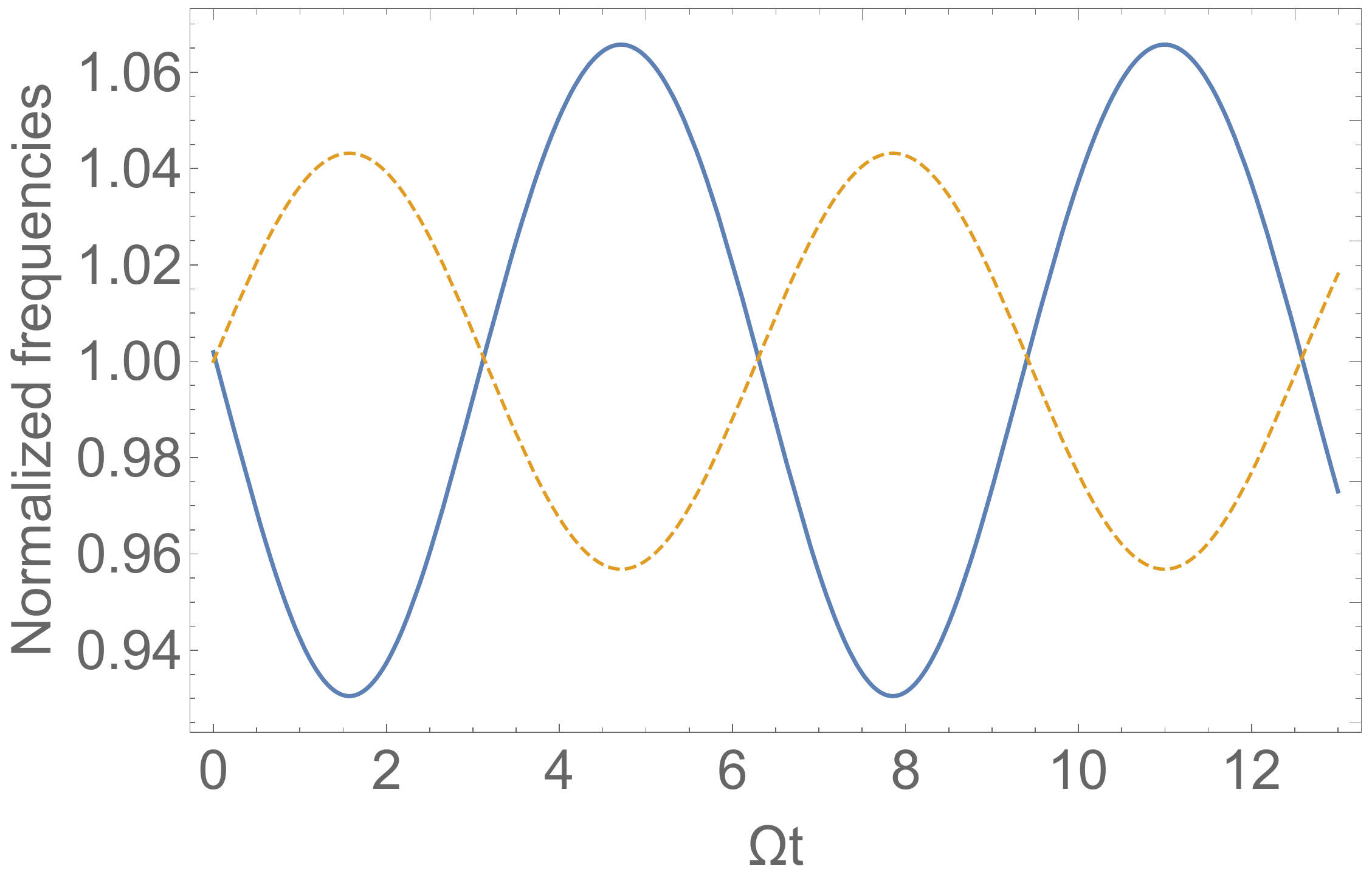}
		\caption{}
	\end{subfigure}
	\begin{subfigure}{0.45\textwidth}
		\includegraphics[width=\linewidth]{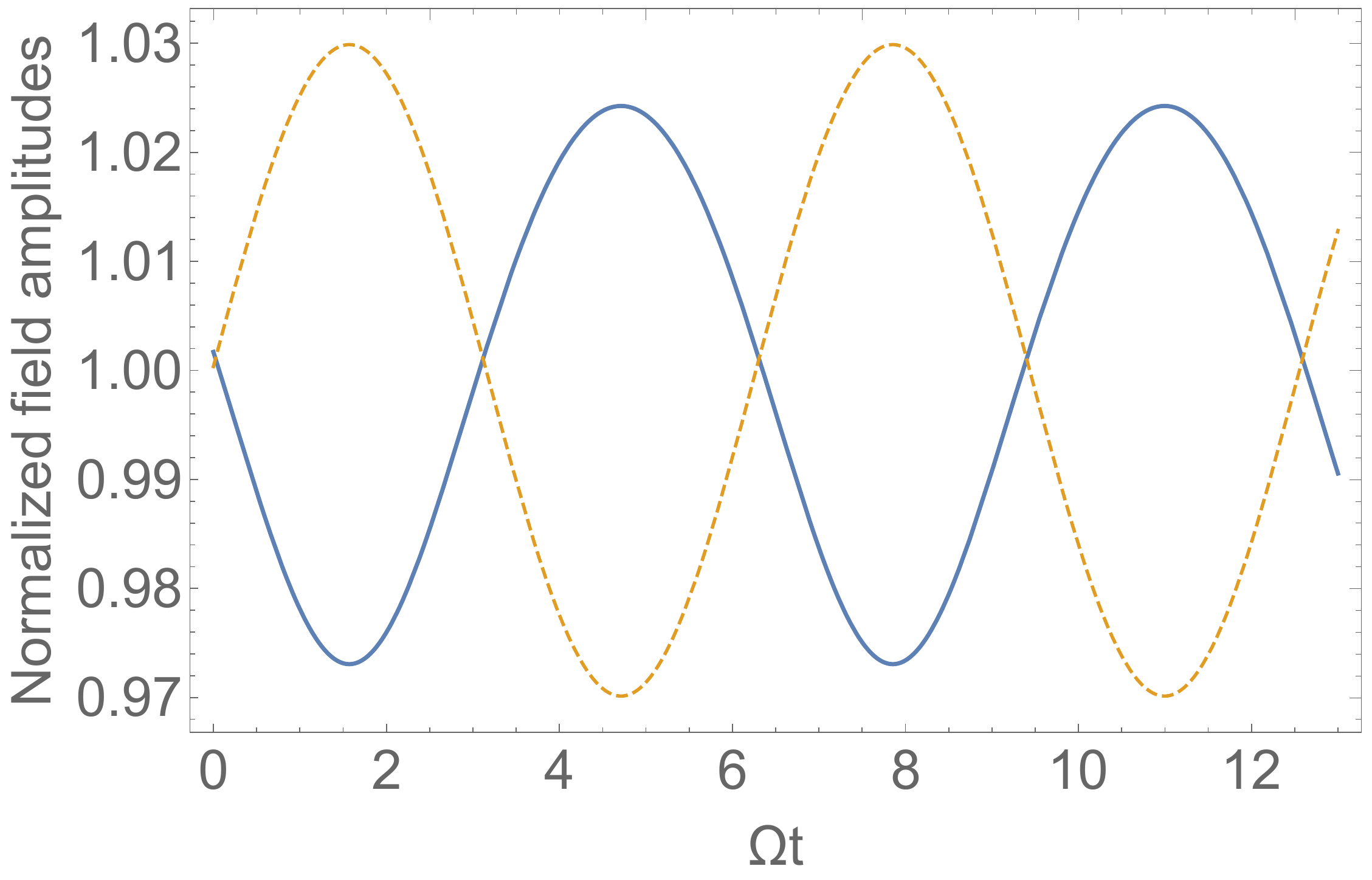}
		\caption{}
	\end{subfigure}
	\caption{(a) Normalized frequencies and (b) normalized field amplitudes of the symmetric (solid line) and antisymmetric (dashed line) cavity modes given by Eqs.~(\ref{norefs})-(\ref{omegakas})  as a function of normalized time $\Omega t$ when the cavity height $d$ is modulated as $d(t) = d_0 (1 + 0.1 \sin(\Omega t))$, where $kd_0 = 1$. Frequencies and field amplitudes are normalized by their time-averaged values.    } 
	\label{Fig:freq&rabifreq}
\end{figure}

\subsection{Field quantization when the cavity thickness is changing
adiabatically}

Let the cavity half-thickness $d$ change with time adiabatically, $d\left(
t\right) $, for given $k$ and $S$. In this case the adiabatic invariant $%
\frac{W_{\mathbf{k,}p}}{\omega _{_{\mathbf{k,}p}}}$ is conserved, where $W_{%
\mathbf{k,}p}$ is an average (observable) energy of the mode. This is
equivalent to conservation of the number of photons in a cavity with slowly
changing parameters. As is well known, the photon number is conserved for a
standard Hamiltonian of an ensemble of harmonic oscillators:

\begin{equation}
\hat{H}=\hbar \sum_{\mathbf{k,}p}\omega _{\mathbf{k,}p}\left( \hat{c}_{%
\mathbf{k,}p}^{\dagger }\hat{c}_{\mathbf{k,}p}+\frac{1}{2}\right),
\label{H of ho}
\end{equation}%
whereas the normalization field is still described by Eqs.~(\ref{symnc}),(%
\ref{asymnc}). At the same time, all results will contain the variables $%
\omega _{_{\mathbf{k,}p}}\left( t\right) $ and $\mathbf{\tilde{E}}_{\mathbf{%
k,}p}$ $\left( t\right) $ which depend on time through their dependence on
the parameter $d(t) $.

%%%%%%%%%%%%%%%%%%%%%%%%%%%%%%

\section{The stochastic equation of evolution for the state vector}

The description of open quantum systems within the stochastic equation of
evolution for the state vector is usually formulated for a Monte-Carlo type
numerical scheme, e.g. the method of quantum jumps \cite{Scully1997,Plenio1998}. We developed an approach suitable for analytic
derivations. Our stochastic equation of evolution is basically the Schr\"{o}dinger equation modified by adding a linear relaxation operator and the
noise source term with appropriate correlation properties. The latter are
related to the parameters of the relaxation operator in such a way that the
expressions for the statistically averaged quantities satisfy certain
physically meaningful conditions.

The protocol of introducing the relaxation operator with a corresponding
noise source term to the quantum dynamics is well known in the Heisenberg
picture, where it is called the Heisenberg-Langevin method \cite{Scully1997,
Gardiner2004, Tokman2013}. Here we use a conceptually similar approach for
the Schr\"{o}dinger equation. The general form of the
stochastic equation of evolution was derived from the Heisenberg-Langevin equations in \cite{tokman2020}. Here we outline how certain physically reasonable constraints on the observables
determine the correlation properties of the noise sources.

\subsection{General properties of the stochastic equation of evolution 
for the state vector}

An open system interacting with a reservoir is generally in a
mixed state and should be described by the density matrix. We are describing
the state of the system with a state vector which has a fluctuating
component. For example, in a certain basis $\left\vert \alpha \right\rangle $
the state vector will be $C_{\alpha }\left( t\right) =\overline{C_{\alpha }}+%
\widetilde{C_{\alpha }}$, where the fluctuating component is denoted with a
wavy bar. The elements of the density matrix of the corresponding mixed
state are $\rho _{\alpha \beta }=\overline{C_{\alpha }C_{\beta }^{\ast }}=%
\overline{C_{\alpha }}\cdot \overline{C_{\beta }^{\ast }}+\overline{%
\widetilde{C_{\alpha }}\cdot \widetilde{C_{\beta }}^{\ast }}$.

The stochastic equation of evolution for the state vector and its Hermitian
conjugate have the general form \cite{tokman2020}
\begin{equation}
\frac{d}{dt}\left\vert \Psi \right\rangle =-\frac{i}{\hbar }\hat{H}%
_{eff}\left\vert \Psi \right\rangle -\frac{i}{\hbar }\left\vert \mathfrak{R}\left(
t\right) \right\rangle  \label{stochastic eq for psi}
\end{equation}
\begin{equation}
\frac{d}{dt}\left\langle \Psi \right\vert =\frac{i}{\hbar }\left\langle \Psi
\right\vert \hat{H}_{eff}^{\dagger }+\frac{i}{\hbar }\left\langle \mathfrak{R}\left(
t\right) \right\vert  \label{hc of stochastic eq for psi},
\end{equation}%
where the non-Hermitian component of the effective Hamiltonian $\hat{H}_{eff}$ corresponds to the relaxation operator and the term  $\left\vert \mathfrak{R}\left( t\right) \right\rangle $ denotes the noise term. We will also
need Eqs.~(\ref{stochastic eq for psi}) and~(\ref{hc of stochastic eq for
psi}) in a particular basis $\left\vert \alpha \right\rangle $:
\begin{equation}
\frac{d}{dt}C_{\alpha }=-\frac{i}{\hbar }\sum_{\nu }\left( \hat{H}%
_{eff}\right) _{\alpha \nu }C_{\nu }-\frac{i}{\hbar }\mathfrak{R}_{\alpha },
\label{c alpha dot}
\end{equation}
\begin{equation}
\frac{d}{dt}C_{\alpha }^{\ast }=\frac{i}{\hbar }\sum_{\nu }C_{\nu }^{\ast
}\left( \hat{H}_{eff}^{\dagger }\right) _{\nu \alpha }+\frac{i}{\hbar }%
\mathfrak{R}_{\alpha }^{\ast },  \label{hc of c alpha dot}
\end{equation}%
where $\mathfrak{R}_{\alpha }=\left\langle \alpha \right. \left\vert \mathfrak{R}\right\rangle $, $%
\left( \hat{H}_{eff}\right) _{\alpha \beta }=\left\langle \alpha \right\vert
\hat{H}_{eff}\left\vert \beta \right\rangle $.

In general, statistical properties of noise that ensure certain
physically meaningful requirements impose certain constraints on the noise
source $\left\vert \mathfrak{R}\right\rangle $ which enters the right-hand side of the
stochastic equation for the state vector. In particular, it is natural to
require that the statistically averaged quantity $\overline{\left\vert
\mathfrak{R}\right\rangle }=0$. We will also require that the noise source $\left\vert
\mathfrak{R}\right\rangle $ has the correlation properties that preserve the norm of
the state vector averaged over the reservoir statistics:
\begin{equation}
\overline{\left\langle \Psi \left( t\right) \right. \left\vert \Psi \left(
t\right) \right\rangle }=1.  \label{stat n of sv}
\end{equation}

\subsection{Noise correlator}

The solution to Eqs.~(\ref{stochastic eq for psi}) and~(\ref{hc of
stochastic eq for psi}) can be formally written as%
\begin{equation}
\left\vert \Psi \right\rangle =e^{-\frac{i}{\hbar }\hat{H}_{eff}t}\left\vert
\Psi _{0}\right\rangle -\frac{i}{\hbar }\int_{0}^{t}e^{\frac{i}{\hbar }\hat{H%
}_{eff}\left( \tau -t\right) }\left\vert \mathfrak{R}\left( \tau \right) \right\rangle
d\tau ,  \label{sol to stochastic eq for psi}
\end{equation}
\begin{equation}
\left\langle \Psi \right\vert =\left\langle \Psi _{0}\right\vert e^{\frac{i}{%
\hbar }\hat{H}_{eff}^{\dagger }t}+\frac{i}{\hbar }\int_{0}^{t}\left\langle 
\mathfrak{R} \left( \tau \right) \right\vert e^{-\frac{i}{\hbar }\hat{H}_{eff}^{\dagger
}\left( \tau -t\right) }d\tau ,  \label{sol to hc of stochastic eq for psi}
\end{equation}%
In the basis $| \alpha \rangle$, Eqs.~(\ref{sol to stochastic eq for psi}),(\ref{sol to hc of
stochastic eq for psi}) can be transformed into%
\begin{equation}
C_{\alpha }=\left\langle \alpha \right\vert e^{-\frac{i}{\hbar }\hat{H}%
_{eff}t}\left\vert \Psi _{0}\right\rangle -\frac{i}{\hbar }%
\int_{0}^{t}\left\langle \alpha \right\vert e^{\frac{i}{\hbar }\hat{H}%
_{eff}\left( \tau -t\right) }\left\vert \mathfrak{R}\left( \tau \right) \right\rangle
d\tau ,  \label{c alpha}
\end{equation}
\begin{equation}
C_{\alpha }^{\ast }=\left\langle \Psi _{0}\right\vert e^{\frac{i}{\hbar }%
\hat{H}_{eff}^{\dagger }t}\left\vert \alpha \right\rangle +\frac{i}{\hbar }%
\int_{0}^{t}\left\langle \mathfrak{R}\left( \tau \right) \right\vert e^{-\frac{i}{\hbar
}\hat{H}_{eff}^{\dagger }\left( \tau -t\right) }\left\vert \alpha
\right\rangle d\tau .  \label{hc of c alpha}
\end{equation}

In order to calculate the observables, we need to know the expressions for
the averaged dyadic combinations of the amplitudes. We can find them using
Eqs.~(\ref{c alpha dot}) and~(\ref{hc of c alpha dot}):%
\begin{eqnarray}
\frac{d}{dt}\overline{C_{\alpha }C_{\beta }^{\ast }} &=&-\frac{i}{\hbar }%
\sum_{\nu }\left( H_{\alpha \nu }^{\left( h\right) }\overline{C_{\nu
}C_{\beta }^{\ast }}-\overline{C_{\alpha }C_{\nu }^{\ast }}H_{\nu \beta
}^{\left( h\right) }\right) -\frac{i}{\hbar }\sum_{\nu }\left( H_{\alpha \nu
}^{\left( ah\right) }\overline{C_{\nu }C_{\beta }^{\ast }}+\overline{%
C_{\alpha }C_{\nu }^{\ast }}H_{\nu \beta }^{\left( ah\right) }\right)  \notag
\\
&&+\left( -\frac{i}{\hbar }\overline{C_{\beta }^{\ast }\mathfrak{R}_{\alpha }}+\frac{i}{%
\hbar }\overline{\mathfrak{R}_{\beta }^{\ast }C_{\alpha }}\right) ,  \label{adc of amp}
\end{eqnarray}%
where we separated the Hermitian and anti-Hermitian components of the
effective Hamiltonian: $\left\langle \alpha \right\vert \hat{H}%
_{eff}\left\vert \beta \right\rangle =H_{\alpha \beta }^{\left( h\right)
}+H_{\alpha \beta }^{\left( ah\right) }$. Substituting Eqs.~(\ref{c alpha})
and~(\ref{hc of c alpha}) into the last term in Eq.~(\ref{adc of amp}), we
obtain%
\begin{eqnarray*}
-\frac{i}{\hbar }\overline{C_{\beta }^{\ast }\mathfrak{R}_{\alpha }}+\frac{i}{\hbar }%
\overline{C_{\alpha }\mathfrak{R}_{\beta }^{\ast }} &=&\frac{1}{\hbar ^{2}}\int_{-t}^{0}%
\overline{\left\langle \mathfrak{R}\left( t+\xi \right) \right\vert e^{-\frac{i}{\hbar }%
\hat{H}_{eff}^{\dagger }\xi }\left\vert \beta \right\rangle \left\langle
\alpha \right. \left\vert \mathfrak{R}\left( t\right) \right\rangle }d\xi \\
&&+\frac{1}{\hbar ^{2}}\int_{-t}^{0}\overline{\left\langle \mathfrak{R}\left( t\right)
\right. \left\vert \beta \right\rangle \left\langle \alpha \right\vert e^{%
\frac{i}{\hbar }\hat{H}_{eff}\xi }\left\vert \mathfrak{R}\left( t+\xi \right)
\right\rangle }d\xi .
\end{eqnarray*}%
To proceed further with analytical results, we need to evaluate these
integrals. The simplest situation is when the noise source terms are
delta-correlated in time (Markovian). In this case only the point $\xi =0$
contributes to the integrals. As a result, Eq.~(\ref{adc of amp})) is
transformed to
\begin{equation}
\frac{d}{dt}\overline{C_{\alpha }C_{\beta }^{\ast }}=-\frac{i}{\hbar }%
\sum_{\nu }\left( H_{\alpha \nu }^{\left( h\right) }\overline{C_{\nu
}C_{\beta }^{\ast }}-\overline{C_{\alpha }C_{\nu }^{\ast }}H_{\nu \beta
}^{\left( h\right) }\right) -\frac{i}{\hbar }\sum_{\nu }\left( H_{\alpha \nu
}^{\left( ah\right) }\overline{C_{\nu }C_{\beta }^{\ast }}+\overline{%
C_{\alpha }C_{\nu }^{\ast }}H_{\nu \beta }^{\left( ah\right) }\right)
+D_{\alpha \beta },  \label{tran adc of amp}
\end{equation}%
where the correlator $D_{\alpha \beta }$ is defined by%
\begin{equation}
\overline{\mathfrak{R}_{\beta }^{\ast }\left( t+\xi \right) \mathfrak{R}_{\alpha }\left( t\right) }%
=\overline{\mathfrak{R}_{\beta }^{\ast }\left( t\right) \mathfrak{R}_{\alpha }\left( t+\xi \right)
}=\hbar ^{2}\delta \left( \xi \right) D_{\alpha \beta }.  \label{cor d}
\end{equation}%
The time derivative of the norm of the state vector is given by%
\begin{equation}
\frac{d}{dt}\sum_{\alpha }\overline{\left\vert C_{\alpha }\right\vert ^{2}}%
=-\sum_{\alpha }\left[ \frac{i}{\hbar }\sum_{\nu }\left( H_{\alpha \nu
}^{\left( ah\right) }\overline{C_{\nu }C_{\alpha }^{\ast }}+\overline{%
C_{\alpha }C_{\nu }^{\ast }}H_{\nu \alpha }^{\left( ah\right) }\right)
-D_{\alpha \alpha }\right] . \label{dot of nor of sv}
\end{equation}%
Clearly, the components $D_{\alpha \alpha }$ of the noise correlator need to
compensate the decrease in the norm due to the anti-Hermitian component of
the effective Hamiltonian. Therefore the expressions for $H_{\alpha \beta
}^{\left( ah\right) }$and $D_{\alpha \alpha }$ have to be mutually
consistent. This is the manifestation of the fluctuation-dissipation theorem
\cite{Landau1965}.

As an example, consider a simple diagonal anti-Hermitian operator $H_{\alpha
\nu }^{\left( ah\right) }$:

\begin{equation}
H_{\alpha \nu }^{\left( ah\right) }=-i\hbar \gamma _{\alpha }\delta _{\alpha
\nu }  \label{ah operator}
\end{equation}
and introduce the following models:

(i) Populations relax much slower than coherences (expected for condensed
matter systems). In this case we can choose $D_{\alpha \neq \beta }=0$, $%
D_{\alpha \alpha }=2\gamma _{\alpha }\overline{\left\vert C_{\alpha
}\right\vert ^{2}}$; within this model the population at each state will be
preserved.

(ii) The state $\alpha =\alpha _{down}$ has a minimal energy, while the
reservoir temperature $T=0$. In this case it is expected that all
populations approach zero in equilibrium whereas the occupation number of
the ground state approaches $1$, similar to the Weisskopf-Wigner model. The
adequate choice of correlators is $D_{\alpha \neq \beta }=0$, $D_{\alpha
\alpha }\propto \delta _{\alpha \alpha _{down}}$, $\gamma _{\alpha
_{down}}=0 $. The expression for the remaining nonzero correlator,%
\begin{equation}
D_{\alpha _{down}\alpha _{down}}=\sum_{\alpha \neq \alpha _{down}}2\gamma
_{\alpha }\overline{\left\vert C_{\alpha }\right\vert ^{2}},
\label{non 0 cor}
\end{equation}
ensures the conservation of the norm:%
\begin{equation*}
\frac{d}{dt}\sum_{\alpha \neq \alpha _{down}}\overline{\left\vert C_{\alpha
}\right\vert ^{2}}=-\sum_{\alpha \neq \alpha _{down}}2\gamma _{\alpha }%
\overline{\left\vert C_{\alpha }\right\vert ^{2}}=-\frac{d}{dt}\overline{%
\left\vert C_{\alpha _{down}}\right\vert ^{2}}.
\end{equation*}%
This is an example of the correlator's dependence on the state vector that
we discussed before.

(iii) A two-level system with states $| 0 \rangle$ and $| 1 \rangle$ and relaxation rates of populations $\frac{1}{T_1}$ and coherence $\frac{1}{T_2} = \frac{1}{2T_1} + \gamma_{el}$, where $\gamma_{el}$ is an elastic relaxation constant. If the equilibrium corresponds to a zero population of the excited state, we have to choose 
$$ \gamma_0 = 0, \; \gamma_1 = \frac{1}{T_2}, \; D_{10} = D_{01} = 0, \; D_{00} = \frac{1}{T_1} \overline{|C_1|^2}, \; D_{11} = 2 \gamma_{el} \overline{|C_1|^2}. $$
It is easy to see that with this choice of relaxation constants and noise correlators Eqs.~(\ref{tran adc of amp}) for $\overline{C_{\alpha }C_{\beta }^{\ast }}$ where $\alpha, \beta = 1,2$ coincide with well-known equations for the density matrix $\rho_{\alpha \beta}$ of a two-level system \cite{Scully1997,fain}. 

%%%%%%%%%%%%%%%%%%%%%%%%

\subsection{Comparison with the Lindblad method}

One can choose the anti-Hermitian Hamiltonian $H_{\alpha \beta }^{\left(
ah\right) }$ and correlators $D_{\alpha \beta }$ in the stochastic equation
of motion in such a way that Eq.~(\ref{tran adc of amp}) for the dyadics $%
\overline{C_{n}C_{m}^{\ast }}$ correspond exactly to the equations for the
density matrix elements in the Lindblad approach. Indeed, the Lindblad form
of the master equation has the form \cite{Scully1997,Plenio1998}%
\begin{equation}
\frac{d}{dt}\hat{\rho}=-\frac{i}{\hbar }\left[ \hat{H},\hat{\rho}\right] +%
\hat{L}\left( \hat{\rho}\right)  \label{Lind mas eq}
\end{equation}%
where $\hat{L}\left( \hat{\rho}\right) $ is the Lindbladian:%
\begin{equation}
\hat{L}\left( \hat{\rho}\right) =-\frac{1}{2}\sum_{k}\gamma _{k}\left( \hat{l%
}_{k}^{\dagger }\hat{l}_{k}\hat{\rho}+\hat{\rho}\hat{l}_{k}^{\dagger }\hat{l}%
_{k}-2\hat{l}_{k}\hat{\rho}\hat{l}_{k}^{\dagger }\right) ,
\label{Lindbladian}
\end{equation}%
Operators $\hat{l}_{k}$ in Eq.~(\ref{Lindbladian}) and their number are
determined by the model which describes the coupling of the dynamical system
to the reservoir. The form of the relaxation operator given by Eq.~(\ref%
{Lindbladian}) preserves automatically the conservation of the trace of the
density matrix, whereas the specific choice of relaxation constants ensures
that the system approaches a proper steady state given by thermal
equilibrium or supported by an incoherent pumping.

Eq.~(\ref{Lind mas eq}) is convenient to represent in a slightly different
form:
\begin{equation}
\frac{d}{dt}\hat{\rho}=-\frac{i}{\hbar }\left( \hat{H}_{eff}\hat{\rho}-\hat{%
\rho}\hat{H}_{eff}^{\dagger }\right) +\delta \hat{L}\left( \hat{\rho}\right)
\label{dot rho}
\end{equation}%
where%
\begin{equation}
\hat{H}_{eff}=\hat{H}-i\hbar \sum_{k}\gamma _{k}\hat{l}_{k}^{\dagger }\hat{l}%
_{k},~~~\delta \hat{L}\left( \hat{\rho}\right) =\sum_{k}\gamma _{k}\hat{l}%
_{k}\hat{\rho}\hat{l}_{k}^{\dagger }.  \label{eff h and delta rho}
\end{equation}%
Writing the anti-Hermitian component of the Hamiltonian in Eqs.~(\ref{c
alpha dot}),(\ref{hc of c alpha dot}) as
\begin{equation}
H_{\alpha \beta }^{\left( ah\right) }=-i\hbar \left\langle \alpha
\right\vert \sum_{k}\gamma _{k}\hat{l}_{k}^{\dagger }\hat{l}_{k}\left\vert
\beta \right\rangle ,  \label{expre for ah operator}
\end{equation}%
and defining the corresponding correlator of the noise source as
\begin{equation}
\overline{\mathfrak{R}_{\beta }^{\ast }\left( t+\xi \right) \mathfrak{R}_{\alpha }\left( t\right) }%
=\hbar ^{2}\delta \left( \xi \right) D_{\alpha \beta },~~~~D_{\alpha \beta
}=\left\langle \alpha \right\vert \delta \hat{L}\left( \hat{\rho}\right)
\left\vert \beta \right\rangle _{\rho _{nm}=\overline{C_{n}C_{m}^{\ast }}},
\label{cor of noise source}
\end{equation}%
we obtain the solution in which averaged over noise statistics dyadics $%
\overline{C_{n}C_{m}^{\ast }}$ correspond exactly to the elements of the
density matrix within the Lindblad method.

Instead of deriving the stochastic equation of evolution of the state vector
from the Heisenberg-Langevin equations we could postulate it from the very
beginning. After that, we could justify the choice of the effective
Hamiltonian and noise correlators by ensuring that they lead to the same
observables as the solution of the density matrix equations with the
relaxation operator in Lindblad form \cite{Plenio1998,blum}. However, the
demonstration of direct connection between the stochastic equation of evolution of the state vector 
and the Heisenberg-Langevin equation provides an important physical insight.

%%%%%%%%%%%%%%%%%%%%%%%%%%

\subsection{Relaxation rates for coupled subystems interacting with a
reservoir}

Whenever we have several coupled subsystems (such as electrons, photon
modes, phonons etc.), each coupled to its reservoir, the
determination of relaxation rates of the whole system becomes nontrivial.
The problem can be solved if we assume that these \textquotedblleft
partial\textquotedblright\ reservoirs are statistically independent.In this
case it is possible to add up partial Lindbladians and obtain the total
effective Hamiltonian.

Consider again the Hamiltonian~(\ref{toh in RWA}) for a two-level
electron system resonantly coupled to two quantized EM cavity modes,
\begin{equation}
\hat{H}=\hbar \omega _{a}\left( t\right) \left( \hat{a}^{\dagger }%
\hat{a}+\frac{1}{2}\right) +\hbar \omega _{b}\left( t\right) \left(
\hat{b}^{\dagger }\hat{b}+\frac{1}{2}\right) +W\hat{\sigma}^{\dagger }\hat{
\sigma} + \hat{V},  
\end{equation}
where 
$$\hat{V} =  - \hat{\sigma}^{\dagger }\left( \chi _{a}\hat{a}+\chi _{b}\hat{b%
}\right)  - \hat{\sigma}\left( \chi _{a}^{\ast }\hat{a}^{\dagger }+\chi
_{b}^{\ast }\hat{b}^{\dagger }\right)  $$
and $\chi _{a,b}\left( t\right) =\mathbf{d\cdot E}_{a,b}$.

Summing up the known (see e.g. \cite{Scully1997,Plenio1998}) partial Lindbladians of
two bosonic (infinite amount of energy levels) and one fermionic (two-level)
subsystems, we obtain
\begin{eqnarray}
L\left( \hat{\rho}\right) &=&-\frac{\gamma }{2}N_{1}^{T_{a}}\left( \hat{%
\sigma}\hat{\sigma}^{\dagger }\hat{\rho}+\hat{\rho}\hat{\sigma}\hat{\sigma}%
^{\dagger }-2\hat{\sigma}^{\dagger }\hat{\rho}\hat{\sigma}\right) -\frac{%
\gamma }{2}N_{0}^{T_{a}}\left( \hat{\sigma}^{\dagger }\hat{\sigma}\hat{\rho}+%
\hat{\rho}\hat{\sigma}^{\dagger }\hat{\sigma}-2\hat{\sigma}\hat{\rho}\hat{%
\sigma}^{\dagger }\right)  \notag \\
&&-\frac{\mu _{a}}{2}\overline{n}_{a}^{T_{em}}\left( \hat{a}\hat{%
a}^{\dagger }\hat{\rho}+\hat{\rho}\hat{a}^{\dagger }\hat{a}-2\hat{a}%
^{\dagger }\hat{\rho}\hat{a}\right) -\frac{\mu _{a }}{2}\left(
\overline{n}_{a}^{T_{em}}+1\right) \left( \hat{a}^{\dagger }\hat{a}
\hat{\rho}+\hat{\rho}\hat{a}\hat{a}^{\dagger }-2\hat{a}\hat{\rho}\hat{a}
^{\dagger }\right)  \notag \\
&&-\frac{\mu _{b }}{2}\overline{n}_{b }^{T_{em}}\left( \hat{b}\hat{b%
}^{\dagger }\hat{\rho}+\hat{\rho}\hat{b}^{\dagger }\hat{b}-2\hat{b}^{\dagger
}\hat{\rho}\hat{b}\right) -\frac{\mu _{b }}{2}\left( \overline{n}%
_{b }^{T_{em}}+1\right) \left( \hat{b}^{\dagger }\hat{b}\hat{\rho}+\hat{%
\rho}\hat{b}\hat{b}^{\dagger }-2\hat{b}\hat{\rho}\hat{b}^{\dagger }\right), 
\label{lindb}
\end{eqnarray}
where $\gamma $ is an inelastic relaxation constant for an isolated atom, $\mu _{a,b}$
are relaxation constants of the EM modes determined by the cavity Q-factor; 
$$
N_{0}^{T_{a}}=\frac{1}{1+e^{-\frac{W}{T_{a}}}}, \; N_{1}^{T_{a}}=\frac{^{e^{-%
\frac{W}{T_{a}}}}}{1+e^{-\frac{W}{T_{a}}}},\; \overline{n}_{a,b}^{T_{em}}=\frac{%
1}{e^{\frac{\hbar \omega _{a,b}}{T_{em}}}-1},
$$
 where $T_{a,em}$ are the
temperatures of the atomic and EM dissipative reservoirs, respectively. It is assumed that these reservoirs are statistically independent.

 For the Lindblad
master equation in the form Eq.~(\ref{dot rho}) we get%
\begin{equation}
\hat{H}_{eff}=\hat{H}-i\hat{\Lambda},  
\label{effectiv ham}
\end{equation}%
where%
\begin{equation}
\hat{\Lambda}=\frac{\hbar }{2}\left\{ \gamma \left( N_{1}^{T_{a}}\hat{\sigma}
\hat{\sigma}^{\dagger }+N_{0}^{T_{a}}\hat{\sigma}^{\dagger }\hat{\sigma}%
\right) 
+ \mu _{a}\left[ \overline{n}_{a }^{T_{em}}\hat{a}\hat{a}%
^{\dagger }+\left( \overline{n}_{a }^{T_{em}}+1\right) \hat{a}^{\dagger
}\hat{a}\right] +\mu _{b}\left[ \overline{n}_{b }^{T_{em}}\hat{b}%
\hat{b}^{\dagger }+\left( \overline{n}_{b }^{T_{em}}+1\right) \hat{b}%
^{\dagger }\hat{b}\right] \right\} .  \label{capital gamma}
\end{equation}%
Using the effective Hamiltonian given by Eqs.~(\ref{effectiv ham}),(\ref
{capital gamma}), we arrive at the stochastic equation for the state vector
in the following form:
\begin{align}
& \left( \frac{\partial }{\partial t}+\gamma _{n_{a}n_{b}1}\right) C_{n_{a}n_{b}1}+i\left(
\left( n_a + \frac{1}{2} \right) \omega _{a}\left( t\right) +\left( n_b +\frac{1}{2} \right) \omega _{b}\left( t\right)
+\frac{W}{\hbar }\right) C_{n_{a}n_{b}1}  \nonumber \\ 
& -  \frac{i}{\hbar }\left\langle n_a \right\vert \left\langle
n_b \right\vert \left\langle 1\right\vert \hat{V}\left\vert \Psi \right\rangle =-
\frac{i}{\hbar }\mathfrak{R}_{n_{a}n_{b}1},  
\label{nanb1}
\end{align}%
\begin{align}
& \left( \frac{\partial }{\partial t}+\gamma _{n_{a}n_{b}0}\right) C_{n_{a}n_{b}0}+i\left(
\left( n_a + \frac{1}{2} \right) \omega _{a}\left( t\right) +\left( n_b +\frac{1}{2} \right) \omega _{b}\left( t\right)
\right) C_{n_{a}n_{b}0}  \nonumber \\ 
& -  \frac{i}{\hbar }\left\langle n_a \right\vert \left\langle
n_b \right\vert \left\langle 0 \right\vert \hat{V}\left\vert \Psi \right\rangle =-
\frac{i}{\hbar }\mathfrak{R}_{n_{a}n_{b}0},  
\label{nanb0}
\end{align}
where
\begin{equation}
\gamma _{n_{a}n_{b}0}=\frac{\gamma }{2}N_{1}^{T_{a}}+\frac{\mu _{a}}{2}\left[
\overline{n}_{a}^{T_{em}}\left( n_{a}+1\right) +\left( \overline{n}%
_{a}^{T_{em}}+1\right) n_{a}\right] +\frac{\mu _{b}}{2}\left[ \overline{n}%
_{b}^{T_{em}}\left( n_{b}+1\right) +\left( \overline{n}_{b}^{T_{em}}+1%
\right) n_{b}\right] ,  \label{gamma0}
\end{equation}
\begin{equation}
\gamma _{n_{a}n_{b}1}=\frac{\gamma }{2}N_{0}^{T_{a}}+\frac{\mu _{a}}{2}\left[
\overline{n}_{a}^{T_{em}}\left( n_{a}+1\right) +\left( \overline{n}%
_{a}^{T_{em}}+1\right) n_{a}\right] +\frac{\mu _{b}}{2}\left[ \overline{n}%
_{b}^{T_{em}}\left( n_{b}+1\right) +\left( \overline{n}_{b}^{T_{em}}+1%
\right) n_{b}\right] ,  \label{gamma1}
\end{equation}%

Eqs.~(\ref{gamma0}),(\ref{gamma1}) determine the rules of
combining the ``partial'' relaxation rates
for several coupled subsystems.

The above expressions include only inelastic relaxation rates. The general procedure of adding elastic relaxation (pure dephasing) is described in \cite{tokman2020}. For the simple RWA models considered in this paper this procedure is reduced to adding $\gamma_{el}$ to $\gamma_{001}$ and changing the noise correlator according to $D_{001;001} \Rightarrow D_{001;001} + 2 \gamma_{el} \overline{|C_{001}|^2}$.

%%%%%%%%%%%%%%%%%%%%%%%%%%%%%

\end{document}